\documentclass[useAMS,usenatbib]{mnras}
\usepackage{newtxtext,newtxmath}
\usepackage[T1]{fontenc}
\usepackage{ae,aecompl}
\usepackage{graphicx}	
\usepackage{amsmath}	
\usepackage{multirow}
\usepackage{lipsum}
\usepackage{blindtext}
\usepackage{xspace}
\usepackage{pdflscape}
\usepackage{afterpage}
\usepackage{gensymb}
\usepackage{booktabs}
\usepackage[usenames, dvipsnames]{color}
\usepackage{lipsum}                     
\usepackage{xargs}                      
\usepackage[pdftex,dvipsnames]{xcolor}  
\usepackage[colorinlistoftodos,prependcaption,textsize=tiny]{todonotes}
\newcommandx{\unsure}[2][1=]{\todo[linecolor=red,backgroundcolor=red!25,bordercolor=red,#1]{#2}}
\newcommandx{\change}[2][1=]{\todo[linecolor=blue,backgroundcolor=blue!25,bordercolor=blue,#1]{#2}}
\newcommandx{\info}[2][1=]{\todo[linecolor=OliveGreen,backgroundcolor=OliveGreen!25,bordercolor=OliveGreen,#1]{#2}}
\newcommandx{\improvement}[2][1=]{\todo[linecolor=Plum,backgroundcolor=Plum!25,bordercolor=Plum,#1]{#2}}
\newcommandx{\thiswillnotshow}[2][1=]{\todo[disable,#1]{#2}}
\def\L{{\mathcal L}}

\title[The impact of jackknife scale on covariances]{\vspace{-0.6cm}Does jackknife scale really matter for accurate large-scale structure covariances?}
\author[Favole et al. 2020]
{\parbox[t]{\textwidth}{\vspace{-0.5cm}Ginevra Favole,$^{1}$\thanks{ginevra.favole@epfl.ch} Benjamin R. Granett,$^2$ Javier Silva Lafaurie,$^3$ Domenico Sapone$^3$}\\
$^1$Institute of Physics, Laboratory of Astrophysics, \'Ecole Polytechnique F\'ed\'erale de Lausanne (EPFL), Observatoire de Sauverny, 1290 Versoix, Switzerland.\\
$^2$INAF-Osservatorio Astronomico di Brera, via Brera 28, 20121 Milano, and via E. Bianchi 46, 23807, Merate, Italy.\\
$^3$Grupo de Cosmolog\'ia y Astrof\'isica Te\'orica, Departamento de F\'{i}sica, FCFM, \mbox{Universidad de Chile}, Blanco Encalada 2008, Santiago, Chile.}
\date{\vspace{-0.4cm}Accepted XXX. Received YYY; in original form ZZZ}
\begin{document}
\label{firstpage}
\pagerange{\pageref{firstpage}--\pageref{lastpage}}
\maketitle
\begin{abstract}
\noindent 
The jackknife method gives an internal covariance estimate for large-scale structure
surveys and allows model-independent errors on cosmological parameters. Using the SDSS-III BOSS CMASS sample, we study how the jackknife size and number of resamplings impact the precision of the covariance estimate on the correlation function multipoles and the error on the inferred baryon acoustic scale. We compare the measurement with the MultiDark Patchy mock galaxy catalogues, and we also validate it against a set of log-normal mocks with the same survey geometry. We build several jackknife configurations that vary in size and number of resamplings.  We introduce the Hartlap factor in the covariance estimate that depends on the number of jackknife resamplings. We also find that it is useful to apply the tapering scheme to estimate the precision matrix from a limited number of resamplings. The results from CMASS and mock catalogues show that the error estimate of the baryon acoustic scale does not depend on the jackknife scale. For the shift parameter $\alpha$, we find an average error of 1.6\%, 2.2\% and 1.2\%, respectively from CMASS, Patchy and log-normal jackknife covariances. Despite these uncertainties fluctuate significantly due to some structural limitations of the jackknife method, our $\alpha$ estimates are in reasonable agreement with published pre-reconstruction analyses. Jackknife methods will provide valuable and complementary covariance estimates for future large-scale structure surveys.
\end{abstract}




\section{Introduction}
\label{sec:intro}
The most popular methods to estimate the uncertainties on the galaxy two-point correlation function (2PCF) internally in a survey are bootstrap \citep{efron1979,davison_hinkley_1997,Norberg2009,Norberg2011} and jackknife \citep{Quenouille1956,Miller1974,Turkey1958,Norberg2009,Norberg2011} resampling. Bootstrap resampling is carried out by randomly selecting $N_{\rm{b}}$ sub-volumes, with replacement, from the original sample. Then the galaxy clustering measurement is performed in each resampling, which is associated a constant weight equal to the number of times that the sub-volume has been selected \citep{Norberg2009}. Similarly, the jackknife method uses $N_{\rm{jk}}$ regions in the survey footprint, each with approximately the same volume. The correlation function is measured on the survey multiple times, each time removing a different jackknife region \citep{Norberg2009}. The covariance matrix is finally inferred from the 2PCF measurements and the 1$\sigma$ errors are derived as the square root of the diagonal elements. 

Internal methods for error estimation are computationally inexpensive and are derived directly from the measurements. Therefore, the analysis does not depend on an assumed cosmological model, which is an attractive feature when testing alternative models such as dark energy or modified gravity. Indeed, jackknife resampling has been widely used to estimate the uncertainties on the galaxy clustering measurements from large-volume spectroscopic surveys \citep[e.g.][]{Zehavi2002,Zehavi2005b,Zehavi2011,Guo2012,Ross2012,2012MNRAS.427.3435A,Guo2015threepoint,Guo2015,Favole2016cameron,Favole2017}. 

Jackknife resampling has two main disadvantages: (i) it tends to overestimate the errors due to the lack of independence between the $N_{\rm{JK}}$ copies; and (ii) it is necessary to balance the number and size of resamplings to be drawn in the survey footprint. The last issue is driven by several factors. First of all, in order to have covariances with reduced noise in their off-diagonal terms, we need a large number of jackknife resamplings. This limits the size of the jackknife regions and also the maximum scales that can be probed in the galaxy clustering observables. It is often assumed in the literature that the jackknife cell size $S_{\rm{JK}}$ should embed the maximum scale measured in the two-point correlation functions. At the same time, to have an invertible (i.e. non-singular) covariance matrix, the number of resamplings should be larger than the number of bins in the measured 2PCF. These conditions are difficult to satisfy in galaxy samples with limited area \citep[e.g.][]{Beutler2011, 2016ApJ...826..154H}. Due to the finite size of any survey footprint, the more resamplings we draw, the smaller their size and the variation between one copy and the next one \citep{Norberg2009,Norberg2011}. 

The issues discussed above have discouraged some cosmologists to use jackknife resampling in favour of estimating the galaxy clustering uncertainties from large sets of independent synthetic galaxy catalogues. In the last years, the advent of efficient codes based on fast gravity solvers has considerably reduced the computational time needed for massive mock production, making available many different realisations of accurate, independent mocks for covariance estimates.
Among these methods, \textsc{PTHALOS} \citep{2002MNRAS.329..629S,2013MNRAS.428.1036M}, \textsc{PINOCCHIO} \citep{2002MNRAS.331..587M,2013MNRAS.433.2389M}, \textsc{PATCHY} \citep{2016MNRAS.456.4156K} and HALOGEN \citep{2015MNRAS.450.1856A} are all based on Lagrangian perturbation theory (LPT). Others, such as QPM \citep{2014MNRAS.437.2594W}, FastPM \citep{2016MNRAS.463.2273F} or PPM-GLAM \citep{2018MNRAS.478.4602K}, use a quick particle mesh approach. Algorithms such as EZ-MOCKS \citep{2015MNRAS.446.2621C} adopt the effective Zel'dovich approximation, while \textsc{COLA} \citep{2013JCAP...06..036T,2016MNRAS.459.2118K}, L-PICOLA \citep{2015A&C....12..109H} or \textsc{ICE-COLA} \citep{2016MNRAS.459.2327I} combine LPT with N-body solvers to speed up the computational time. Finally, high-fidelity mocks can also be obtained from multiple realisations of a log-normal density field \citep{Coles1991,Beutler2011,2016ApJ...826..154H,2017JCAP...10..003A,2019MNRAS.482.1786L}, or populating dark matter haloes with galaxies using the halo occupation distribution technique \citep[HOD; e.g.][]{2002ApJ...575..587B,2005ApJ...633..791Z}.

All these fast mocking approaches are extremely convenient compared to running a full N-body code, but they are generally limited to predicting the dark matter distribution. On top of the dark matter field, it is necessary to model the galaxy distribution by properly accounting for the different baryonic components and the complex physics of galaxy formation and evolution \citep[e.g.][]{2010gfe..book.....M,doi:10.1146/annurev-astro-081913-040019}. 

Simulating baryons is a non-trivial task, which requires advanced computational techniques and resources. Semi-analytic models of galaxy formation and evolution \citep[SAMs;][]{1991ApJ...379...52W,1993MNRAS.264..201K,2006RPPh...69.3101B,2010PhR...495...33B,2012NewA...17..175B,doi:10.1146/annurev-astro-082812-140951,2006MNRAS.368.1540C,2018MNRAS.479....2C, Croton2006,2016ApJS..222...22C,2015MNRAS.446.3820G,2018MNRAS.481..954C} and hydrodynamical simulations \citep[e.g.][]{1997MNRAS.284..235Y,2003MNRAS.339..312S,2003ApJ...598...73Y,2010MNRAS.401..791S,2014Natur.509..177V,2014MNRAS.444.1518V,2014MNRAS.445..175G, 2015MNRAS.450.1937C,2015MNRAS.446..521S,2018MNRAS.473.4077P} are now able, with different degrees of accuracy, to incorporate the multitude of ingredients and physical processes that contribute to shape the formation and evolution of galaxies within their host dark matter haloes. Some of these processes are: gas accretion \citep{guo11,henriques13,hirschmann16} and cooling \citep{10.1111/j.1365-2966.2010.16806.x,monaco14,hou17}, star formation \citep{lagos11}, stellar winds \citep{lagos13}, stellar evolution \citet{tonini09,henriques11,gp14}, AGN feedback \citet{Bower2006,Croton2006} or environmental processes \citet{Weinmann2006,font08,2017MNRAS.471..447S,2018MNRAS.481..954C}.

Analogously, one should also account for the effect of massive neutrinos on the growth of cold dark matter perturbations, which are responsible of suppressing the matter power spectrum at intermediate and small scales \citep{2013MNRAS.428.3375A,Wright_2017,2019JCAP...01..010P}. 

All of these assumptions and prescriptions have uncertainties which become significant on non-linear scales and limit the accuracy of the covariance estimate. 

Upcoming surveys, such as the Dark Energy Spectroscopic Instrument\footnote{\url{https://www.desi.lbl.gov}} \citep[DESI;][]{2015AAS...22533607S}, Euclid\footnote{\url{https://www.euclid-ec.org}} \citep{2011arXiv1110.3193L,2016MNRAS.459.1764S} and the Large Synoptic Survey Telescope\footnote{\url{https://www.lsst.org}} \citep[LSST;][]{2019ApJ...873..111I}, will bring us to the era of high precision cosmology. In order to prepare to this new phase, it is imperative to improve and compare different methods to construct accurate covariances able to capture the hidden physical process of gravitational collapse. These methods have to carefully optimise the specific binning scheme adopted in order to minimise the noise in the measurements.

For the reasons above, in this work we aim to rehabilitate the use of jackknife resamplings versus mocks for estimating covariances. We explore how varying the size ($S_{\rm{JK}}$) and number ($N_{\rm{JK}}$) of jackknife regions impacts the precision in the error estimates of galaxy clustering and on the baryon acoustic oscillation scale through the $\alpha$ shift parameter. In concrete, we measure the monopole and quadrupole two-point correlation functions of the BOSS CMASS DR12 galaxies and we compute their covariances using four different jackknife configurations, coupled with two binning schemes. We compare these results with those obtained from 1000 MultiDark Patchy mocks implemented for BOSS DR12 \citep{2016MNRAS.456.4156K} and with the average covariance obtained by performing jackknife resampling on 10 Patchy mocks, randomly chosen among the 1000. We then contrast these results with 1000 independent log-normal mocks with the same volume of CMASS, and we also perform the jackknife analysis on 10 of the log-normal mock realisations.

From these covariances, we build the precision matrices needed to estimate the $\alpha$ shift parameter through a Monte Carlo Markov Chain (MCMC) algorithm.  We reduce the noise in their off-diagonal terms by applying the tapering correction \citep{Kaufman:2008}. We study the impact of a variation in the tapering parameter, $T_{\rm{p}}$, on the $\alpha$ results.
These estimates of $\alpha$ will be directly compared with the galaxy clustering pre-reconstruction results from the BOSS collaboration \citep{10.1093/mnras/stw2372}.

The paper is organised as follows: in Sec.\,\ref{sec:data}, we introduce the observational galaxy sample used in our analysis, SDSS-III/BOSS CMASS DR12; in Sec.\,\ref{sec:measurements}, we present the galaxy clustering measurements performed, together with the jackknife methodology and schemes used to estimate their uncertainties. Sec.\,\ref{sec:models} describes the models adopted to analyse the CMASS observations: the MultiDark Patchy mocks (\S\,\ref{sec:patchy}), the log-normal ones (\S\,\ref{sec:mocks}), and the analytic models constructed for the Monte Carlo runs (\S\,\ref{sec:analyticmodel}). In Sec.\,\ref{sec:javiermethod}, we explain the Monte Carlo algorithm used to extract the $\alpha$ BAO parameter. Sec.\,\ref{sec:results} presents our main results, which are discussed in Sec.\,\ref{sec:discussion}, together with our conclusions.

Throughout the paper we adopt a \citet{Planck2014} cosmology with $\Omega_m=0.307115$, $\Omega_{\Lambda}=0.692885$, $h=0.6777$, $n=0.9611$ and $\sigma_8=0.8228$. 

\section{Observed galaxy sample: BOSS CMASS DR12}
\label{sec:data}

The SDSS-III BOSS survey observed about 1.2 million galaxies over an effective area of 9329\,deg$^2$ \citep[]{2017MNRAS.470.2617A}, using the 2.5m Sloan Telescope \citep{Gunn2006} at the Apache Point Observatory in New Mexico. It used a drift-scanning mosaic CCD camera with five photometric bands, $ugriz$ \citep[][]{Gunn1998, 1996AJ....111.1748F}, and two spectrographs covering the wavelength range $3600-10000{\buildrel _{\circ} \over {\mathrm{A}}}$ with a resolving power of 1500 to 2600 \citep[]{Smee2013}. Spectroscopic redshifts were measured using the minimum-$\chi^2$ template-fitting procedure by \cite{Aihara2011}, with templates from \cite{2012AJ....144..144B}.

BOSS targeted galaxies into two main samples: LOWZ at $z<0.43$ and CMASS at $0.43<z<0.7$ \citep[][]{2012ApJS..203...21A}. For our analysis, we use the data from the BOSS CMASS DR12 sample \citep[][]{2015ApJS..219...12A, 2016MNRAS.455.1553R, 2017MNRAS.470.2617A}, which is defined through a number of magnitude and colour cuts aimed at obtaining a selection of galaxies with approximately constant stellar mass \citep[see e.g.][]{2016MNRAS.455.1553R}.


\section{Measurements}
\label{sec:measurements}
\subsection{Two-point correlation functions}
\label{sec:2pt}
We measure the two-point correlation function, $\xi(s,\mu)$, of the galaxy sample described in Sec.\,\ref{sec:data} using the code from \citet{Favole2017}. This is based on the Landy-Szalay estimator \citep{1993ApJ...412...64L},
\begin{equation}
    \xi(s,\mu)=\frac{DD(s,\mu)-2DR(s,\mu)+RR(s,\mu)}{RR(s,\mu)},
    \label{eq:landy}
\end{equation}
where $s$ is the redshift-space distance and $\mu$ is the cosine of the angle between $s$ and the line of sight. 

In the expression above, $DD$, $DR$ and $RR$ are respectively the data-data, data-random and random-random pair counts that we can form between the galaxy and random catalogues. The latter is built to have the same angular footprint and radial distribution of the CMASS observations. All the pairs above are properly normalised by the number of galaxies (randoms) and weighted to correct from different systematic effects \citep[see e.g.][]{2012MNRAS.425..415S,Ross2012, Favole2016b}. In particular, we weight the observed data for potential fibre collisions ($w_{\rm{fc}}$) and for redshift failures ($w_{\rm{zf}}$). We also account for possible variation in the galaxy (random) number densities assuming the FKP \citep[][]{1994ApJ...426...23F} weight:
\begin{equation}
    w_{\rm{FKP}}=\frac{1}{1+n(z)P_{0}},
    \end{equation}
where $n(z)$ is the galaxy (random) number density at redshift $z$ and $P_0$ is a constant quantity that roughly corresponds to the amplitude of the CMASS power spectrum at $k=0.1\,h$Mpc$^{-1}$. We assume $P_0=20000\,h^3$Mpc$^{-3}$ as in \citet[][]{2012MNRAS.427.3435A}. 

From Eq.\,\ref{eq:landy}, we compute the multipoles of the CMASS correlation function as:
\begin{equation}
    \xi_l(s)=\frac{2l+1}{2}\int^{1}_{-1}\xi(s,\mu)P_l(\mu)\,d\mu,
\end{equation}
where $P_l(\mu)$ are the Legendre polynomials. In this study, we focus only on the first two even multipoles of the 2PCF, i.e. the monopole $\xi_0(s)$ and the quadrupole $\xi_2(s)$. We explore two different binning schemes, both centered on the BAO distance scale, coupled with the jackknife configurations defined in Sec.\,\ref{sec:jkconfig}: (i) 20 linear bins in $24<s<184\,h^{-1}$Mpc and 120 linear bins in $0<\mu<1$; (ii) 10 linear bins in $24<s<184\,h^{-1}$Mpc and 120 linear bins in $0<\mu<1$.


\subsection{Jackknife configurations and covariances}\label{sec:jkconfig}
We implement jackknife resampling in the BOSS CMASS DR12 galaxy sample following a standard prescription \citep[see e.g.][]{Norberg2009,Norberg2011} and adopting four different configurations, which are summarised in Table\,\ref{tab:jkconfig}. We divide the survey footprint into 200, 100, 50 and 20 RA$\times$DEC cells approximately containing the same number of galaxies (randoms). Since the bias changes with redshift, we choose not to vary the size of the jackknife regions in the line-of-sight direction, as they would not be representative of the clustering signal.
The CMASS covariance matrix for $N_{\rm{JK}}$ jackknife resamplings is \citep[e.g.][]{Ross2012,Favole2016cameron}: 
\begin{equation} 
  C_{\rm{ij}}=\frac{N_{\rm{JK}}-1}{N_{\rm{JK}}}\sum_{\rm{a=1}}^{N_{\rm{JK}}} (\xi^a_{\rm{i}}-\bar{\xi}_{\rm{i}})(\xi^a_{\rm{j}}-\bar{\xi}_{\rm{j}}),
  \label{eq:covma}
\end{equation}
where $\bar{\xi}_{i}$ is the mean jackknife correlation function in the $i^{\rm{th}}$ bin,
\begin{equation} 
  \bar{\xi}_{\rm{i}}=\sum_{\rm{a=1}}^{N_{\rm{JK}}} \xi^a_{\rm{i}}/N_{\rm{JK}}.
\end{equation}
The overall factor in Eq.\;\ref{eq:covma} corrects from the lack of independence
between the $N_{\rm{JK}}$ jackknife copies, which is the main limitation of the jackknife method. In fact, from one configuration to the next, $N_{\rm{JK}}-2$ cells are the same \citep{Norberg2011}. 

\begin{table}\centering
\begin{tabular}{@{}lcc@{}}
\toprule
$N_{\rm{JK}}$& $A_{\rm{JK}}$\,\,[deg$^2$]&$S_{\rm{JK}}$\,\,[$h^{-1}$Mpc] \\
\midrule
200&46.6&110.7\\
100&93.3&156.6\\
50&186.6&221.4\\
20&932.9&495.1\\
\bottomrule
\end{tabular}
\caption{Jackknife configurations adopted in our analysis. For each of the four cases implemented, we indicate the number of jackknife resamplings ($N_{\rm{JK}}$), the area ($A_{\rm{JK}}$) and comoving size ($S_{\rm{JK}}$) of the individual cell computed in \citet{Planck2014} cosmology at the mean redshift of CMASS, $z=0.56$.}
\label{tab:jkconfig}
\end{table}


\section{Models}
\label{sec:models}
\subsection{MultiDark Patchy mock galaxy catalogues}
\label{sec:patchy}
In order to analyse the CMASS two-point correlation function multipoles, we use 1000 MultiDark Patchy mocks implemented for BOSS DR12 \citep{2016MNRAS.456.4156K,2016MNRAS.460.1173R}. These were constructed using the Patchy code \citep{2014MNRAS.439L..21K}, which couples second-order Lagrangian perturbation theory \citep[2LPT; see e.g.][]{1994MNRAS.267..811B, 1995A&A...296..575B, 1995MNRAS.276...39C} with a spherical collapse model \citep{1994ApJ...427...51B,2006MNRAS.365..939M,2013MNRAS.428..141N} to generate a dark matter field on a mesh \citep{2013MNRAS.435L..78K}. 
The mesh has been populated with galaxies using a deterministic and a stochastic bias, whose parameters were constrained to precisely match the two- and three-point statistics \citep{2015MNRAS.450.1836K}.

The Patchy mocks were built assuming the same $\Lambda$CDM Planck cosmology adopted in this work (see \S\,\ref{sec:intro}), and they were calibrated against N-body-based reference catalogues through subhalo abundance matching to recover the survey geometry, selection function, galaxy number density and clustering bias of BOSS data in 10 redshift bins \citep{2016MNRAS.460.1173R}. Previous studies \citep[e.g.][]{2015MNRAS.452..686C,2016MNRAS.456.4156K,2016MNRAS.460.1173R} have demonstrated that the Patchy mocks are very accurate in reproducing the two- and three-point statistics of BOSS data.

\subsection{Log-normal mock galaxy catalogues}
\label{sec:mocks}
Besides the Patchy mocks described in the previous section, we perform the analysis also using a set of 1000 log-normal mock galaxy catalogues that we generate for the BOSS CMASS sample at mean redshift $z\sim0.56$. In this way, we validate the accuracy of our log-normal mock algorithm, which is a useful tool for future studies, with a much lower computational cost compared to the Patchy method.

The target power spectrum for our LN mocks was computed with the code \emph{CLASS}\footnote{\url{https://github.com/lesgourg/class_public}} \citep{Blas2011}, assuming the formalism used for our analytic models and given in Eq.\,\ref{eq:Pk_non_lin} We have then applied a linear
bias and the Halofit \citep{Takahashi2012} prescription to model the non-linear galaxy power spectrum: $P_{\rm nl}(k) = b^2 P(k) + c k^{-3}$ with the value $b=1.87$ and $c=0.021$. The $k^{-3}$ term is used to match the amplitude of the correlation function on BAO scales and is motivated by the similar nuisance parameters introduced in the BAO fitting procedure (Eq. \ref{eq:modelxi0}). The free parameters in the model were tuned to match the measured CMASS monopole correlation function. 

We present the \emph{Synmock} code used to produce the log-normal catalogues in a public repository\footnote{\url{https://github.com/bengranett/synmock}}. The implementation follows the standard approach for generating log-normal simulations \citep[see also][]{Beutler2011,Pearson2016}. For each realization, first we generated a Gaussian density field $\delta_G(\vec{x})$ on a cubic grid with dimension
$L=4096 h^{-1}$Mpc and step size $h=4 h^{-1}$Mpc and transformed it to derive
the target log-normal field: 
\begin{equation}
\delta(\vec{x}) = \exp\left({\delta_G(\vec{x}) - \sigma^2/2}\right) - 1\,,
\end{equation}
where $\sigma^2$ is the variance of the Gaussian field. In order to match the target power spectrum we made a Fourier transform of the power spectrum to compute the correlation function and using the relationship \citep{Coles1991} 
\begin{equation}
\xi_G(|\vec{x}-
\vec{x}'|) = \log\left(1+\xi(|\vec{x}-\vec{x}'|)\right)\,.
\end{equation}

The log-normal density field was used to build a discrete galaxy field by Poisson sampling the number density
$n(\vec{x})=\bar{n}(1+\delta(\vec{x}))$. We applied a uniform random offset to move the mock galaxies away from the grid points.

The velocity field was computed on the same $\delta(x)$ grid using the linear continuity equation in Fourier space:
\begin{equation}
\vec{v}\big(\vec{k}\big) = i\frac{f a H}{b}\delta\big(\vec{k}\big) \frac{\vec{k}}{k^2}\,, 
\end{equation}
where $f$ is the logarithmic growth rate. After a final Fourier transform, the velocity of each galaxy was assigned using the value at the nearest grid point. 

We built 1000 BOSS CMASS realizations (\textquoteleft\textquoteleft LN mocks", hereafter) with $0.43<z<0.7$ by cutting the BOSS survey geometry in the log-normal simulation box above. The Cartesian galaxy coordinates were transformed to the spherical coordinates right ascension, declination and radial distance with the origin at the center of the simulation box.  In order to transform to the redshift-space coordinates, the line-of-sight peculiar velocity component was computed and applied to the radial comoving distance: $r_s = r + \vec{r}\cdot\vec{v}/\left(a H |\vec{r}|\right)$.  We constructed a coarse angular mask using the Healpix \citep{Gorski2005} scheme at resolution $n_{\rm{side}}=64$ and discarded galaxies outside the mask. The catalog was further downsampled along the radial direction to match the target redshift distribution.  We generated an unclustered random catalogue with 10 times the number density of the CMASS data that precisely corresponds to the mock construction using the same angular mask and radial selection function.

After computing the correlation functions of the  {\bf $N_{\rm{LN}}=1000$} log-normal mocks, we derive their covariance matrix as:
\begin{equation} 
  C_{\rm{ij}}=\frac{1}{N_{\rm{LN}}-1}\sum_{\rm{a=1}}^{N_{\rm{LN}}} (\xi^a_{\rm{i}}-\bar{\xi}_{\rm{i}})(\xi^a_{\rm{j}}-\bar{\xi}_{\rm{j}}),
  \label{eq:covmocks}
\end{equation}
where $\bar{\xi}_{\rm{i}}$ is their mean 2PCF in the $i^{\rm{th}}$ bin. The pre-factor properly accounts for the fact that the mock realisations are  independent.


\subsection{Analytic models}
\label{sec:analyticmodel}
Besides the Patchy and the LN mocks, we also model the multipoles of the BOSS CMASS two-point correlation function using an analytic approach, which is required to run the Monte Carlo analysis (see Sec.\,\ref{sec:javiermethod}). The 2PCF can be obtained from the Fourier transform of the matter power spectrum, $P(k)$, for which we assume the template from~\cite{Padmanabhan:2008ag}:
\begin{equation}
\label{eq:Pk_non_lin}   
    P(k)=\left[P_{\rm{lin}}(k)-P_{\rm{dw}}(k)\right]e^{-k^2\Sigma_{\rm{nl}}^2/2}+P_{\rm{dw}}(k)\,.
\end{equation}
In the equation above, $P_{\rm{lin}}(k)$ is the linear matter power spectrum computed using the Boltzmann code CLASS~\citep{Lesgourgues:2011re} assuming the Planck 2015~\citep{Ade:2015xua} fiducial cosmology. The $P_{dw}(k)$ term is the de-wiggled power spectrum~\citep{Eisenstein:1997ik}, while the $\Sigma_{nl}$ parameter encodes the smoothing of the BAO peak due to non-linear effects~\citep{Crocce:2005xy}. The multipoles of the analytic 2PCF are defined as:
\begin{equation}
\label{eq:multipoles_xi}
    \xi_l(s) = \frac{i^l}{2\pi^2}\int_0^{\infty} P_l(k)j_l(ks)k^2dk\,,
\end{equation}
from which we recover the monopole ($l=0$) and the quadrupole ($l=2$). In Eq.\,\ref{eq:multipoles_xi}, $j_l(x)$ represents the spherical Bessel function of first kind and order $l$, while $P_l(k)$ are the multipoles of the power spectrum defined as:
\begin{equation}
\label{eq:multipoles_pk}
    P_l(k)=\frac{2l+1}{2}\int^1_{-1}\left(1+f\mu^2\right)^2P(k)L_l(\mu)d\mu\,,
\end{equation}
where $L_l(x)$ is the Legendre polynomial of order $l$ and $P(k)$ is the template given in Eq.\,\ref{eq:Pk_non_lin}. 
By replacing Eq.\,\ref{eq:multipoles_pk} in Eq.\,\ref{eq:multipoles_xi}, the analytic expressions for monopole ($l=0$) and quadrupole ($l=2$) are respectively~\citep{2012MNRAS.427.2146X}:
\begin{equation}
\label{eq:modelxi0}
 \xi_{\rm{model}}^{(0)}(s) = B_0\xi_0(\alpha s)+a_0^{(0)}+\frac{a_1^{(0)}}{s}+\frac{a_2^{(0)}}{s^2}\,,  
\end{equation}
\begin{equation}
\label{eq:modelxi2}
 \xi_{\rm{model}}^{(2)}(s) = B_2\xi_2(\alpha s)+a_0^{(2)}+\frac{a_1^{(2)}}{s}+\frac{a_2^{(2)}}{s^2}\,,  
\end{equation}
where $\alpha$ is the shift parameter, while $(a_1^{(i)},a_2^{(i)},a_3^{(i)})$ are linear nuisance parameters. 

The shift parameter $\alpha$ in Eqs.\,\ref{eq:modelxi0} and \ref{eq:modelxi2} is usually defined as~\citep{Padmanabhan:2008ag}:
\begin{equation}
\label{shift_parameter}
\alpha = \frac{D_V}{r_s}\frac{r_s^{\rm{fid}}}{D_V^{\rm{fid}}}\,,
\end{equation} 
where $r_s$ represents the sound horizon \citep{Hu:1995en}, and $D_V$ the volume-averaged distance given by \citep{Eisenstein:2005su}:
\begin{equation}
    D_{\rm V}(z)=\left[cz(1+z)^2D_{\rm A}^2(z)H^{-1}(z)\right]^{1/3}, 
\end{equation}
with  $D_{\rm A}(z)$ being the angular diameter distance, and $H(z)$ the Hubble parameter at redshift $z$.
The $\alpha$ shift parameter accounts for the observed distortion between distances due to the chosen fiducial cosmology, while the nuisance parameters $(a_1^{(i)},a_2^{(i)},a_3^{(i)})$ and $B_1,B_2$ incorporate those effects that are responsible of modulating the clustering amplitude, such as redshift-space distortions~\citep{2012MNRAS.427.2146X}, linear bias, and the power spectrum normalisation, $\sigma_8$.


\section{shift parameter estimation}
\label{sec:javiermethod}
Following the methodology presented in \citet{PRL}, we analyse the BOSS CMASS covariances from jackknife, and the Patchy and LN ones with and without jackknife, using a Monte Carlo Markov Chain based on a Metropolis-Hastings algorithm\footnote{\url{https://emcee.readthedocs.io/en/stable/}}. Our MCMC code is publicly available on GitHub\footnote{\url{https://github.com/javiersilvalafaurie/BTCosmo}}.

In order to find the optimal parameter values, we assume a likelihood function $\L\propto\exp(-\chi^2/2)$, with 
\begin{equation}
\chi^2 = \left(\vec{\xi}_{\rm{model}}-\vec{\xi}_{\rm{obs}}\right)^T\,\hat{\Psi}\,\left(\vec{\xi}_{\rm{model}}-\vec{\xi}_{\rm{obs}}\right),
\label{eq:chi2}
\end{equation}
where 
\begin{equation}
\vec{\xi}_{\rm{model}}\equiv\left(\vec{\xi}^{(0)}_{\rm{model}},\vec{\xi}^{(2)}_{\rm{model}}\right)
\end{equation}
represents the theoretical correlation function whose components are given in Eqs.~\ref{eq:modelxi0}-\ref{eq:modelxi2}, while $\vec{\xi}_{\rm{obs}}$ corresponds to the observed CMASS 2PCF, with its monopole and quadrupole  moments grouped in a vector as a function of the comoving distance:
\begin{equation}
\vec{\xi}_{\rm{obs}}\equiv\left(\vec{\xi}^{(0)}_{\rm{CMASS}},\vec{\xi}^{(2)}_{\rm{CMASS}}\right).
\end{equation}
The $\Psi$ term in Eq.\,\ref{eq:chi2} is the precision matrix defined as: 
\begin{equation}
\label{final_pres_matrix}
    \hat{\Psi} = \left(1-\frac{n_{\rm{b}}+1}{N_{\rm{res}}-1}\right)\left(\hat{C}\circ T\right)^{-1}\circ T\,,
\end{equation}
where $\hat{C}$ is the total assembled covariance matrix from the individual auto- and cross-covariances either obtained from 1000 mock realisations of from jackknife resamplings:
\begin{equation}
\label{cov_matrix_0_2}
\hat{C} = 
\begin{pmatrix}
    \hat{C}_{\xi_0\xi_0} & \hat{C}_{\xi_0\xi_2} \\\\
    \hat{C}_{\xi_0\xi_2}^T & \hat{C}_{\xi_2\xi_2}\\
\end{pmatrix}\,.
\end{equation}
The first term in parenthesis in Eq.\,\ref{final_pres_matrix} is the Hartlap correction factor \citep{Hartlap:2006kj}, which corrects from the bias introduced in the covariance matrix by the limited number of jackknife resamplings and 2PCF bins. In Tab.\,\ref{tab:hartlap}, we report the values of the Hartlap factor as a function of the number of jackknife resamplings and bins used in our analysis. In the last row of the table, we also show the value of the Hartlap correction for 20 spatial bins and 1000 mock realisations, i.e. our  ideal 1000 Patchy/LN configurations. In these cases the Hartlap correction is almost negligible, as it is very close to unity. The lower the number of bins and jackknife resamplings, the stronger the effect of such  a correction. 

The quantity $T$ in Eq.\,\ref{final_pres_matrix} is the tapering correction \citep{Kaufman:2008} that minimises the noise in the off-diagonal terms of the covariance matrix; for further details see also \cite{PRL}. In this work, we assume a tapering parameter $T_{\rm{p}}=500$\,h$^{-1}$Mpc to ensure that the entire covariance matrix is positive semi-definite and the deviations in the off-diagonal terms are minimised. In Sec.\,\ref{sec:results}, we test how a variation in the tapering parameter affects the results for $\alpha$ and its uncertainty. Further details on the dependence of $\alpha$ on $T_{\rm{p}}$ are addressed also in \citet{2015MNRAS.454.4326P}. 
\begin{table}\centering
\begin{tabular}{@{}lcc@{}}
\toprule
$n_{\rm{b}}$& $N_{\rm{JK}}$ &Hartlap correction \\
\midrule
10&20&0.42105\\
20&50&0.57143\\
20&100&0.78788\\
20&200&0.89447\\
\midrule
 20&1000&0.97898\\
\bottomrule
\end{tabular}
\caption{Values of the Hartlap factor \citep{Hartlap:2006kj} as a function of the number of bins $n_{\rm{b}}$ and jackknife resamplings $N_{\rm{JK}}$ used in our analysis. In the last row we also show the value of the Hartlap correction for our ideal 1000 Patchy/LN mocks. }
\label{tab:hartlap}
\end{table} 

\section{Results}

\label{sec:results}
\begin{figure}
\begin{center}
\includegraphics[width=\linewidth]{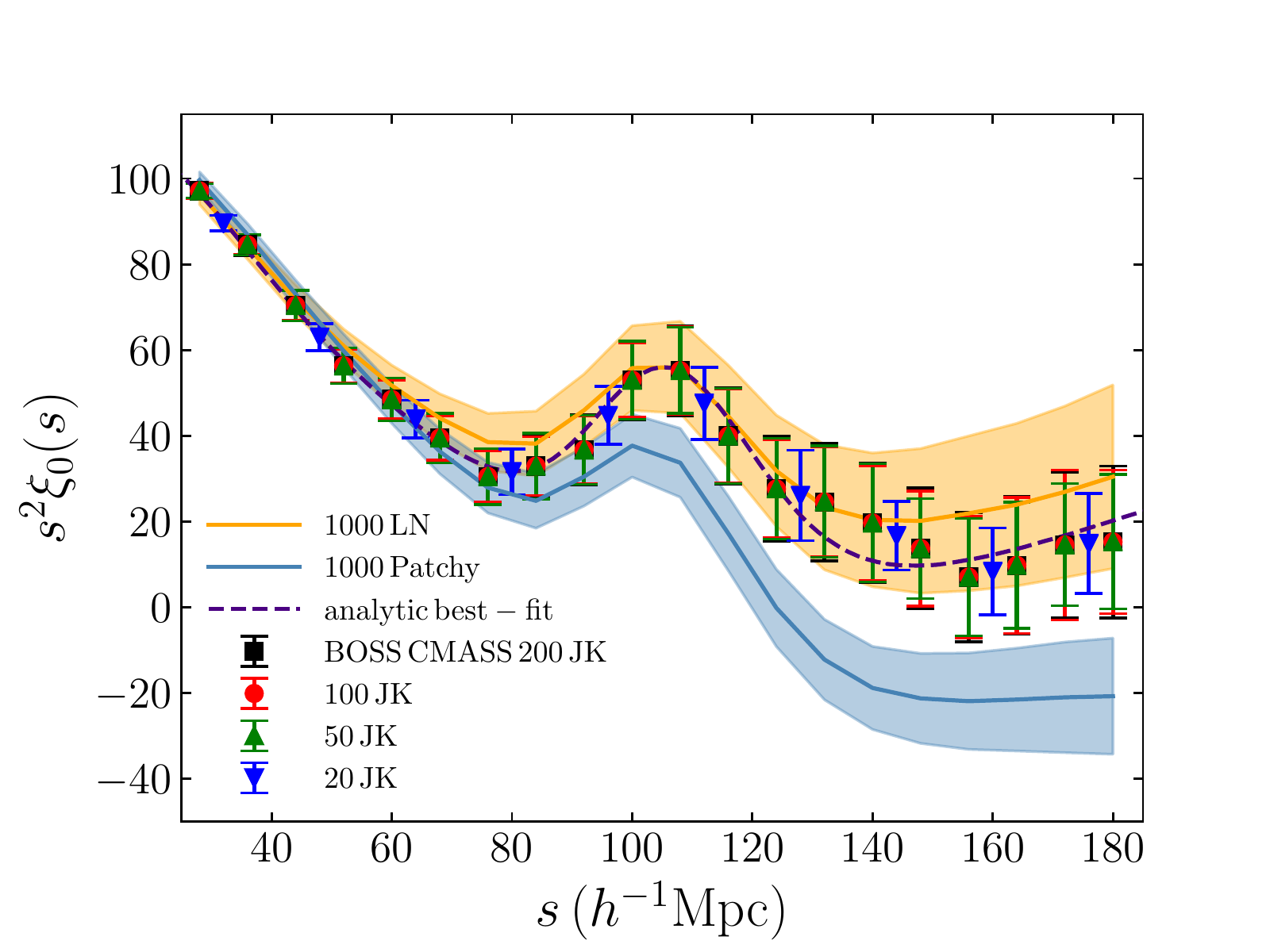}
\includegraphics[width=\linewidth]{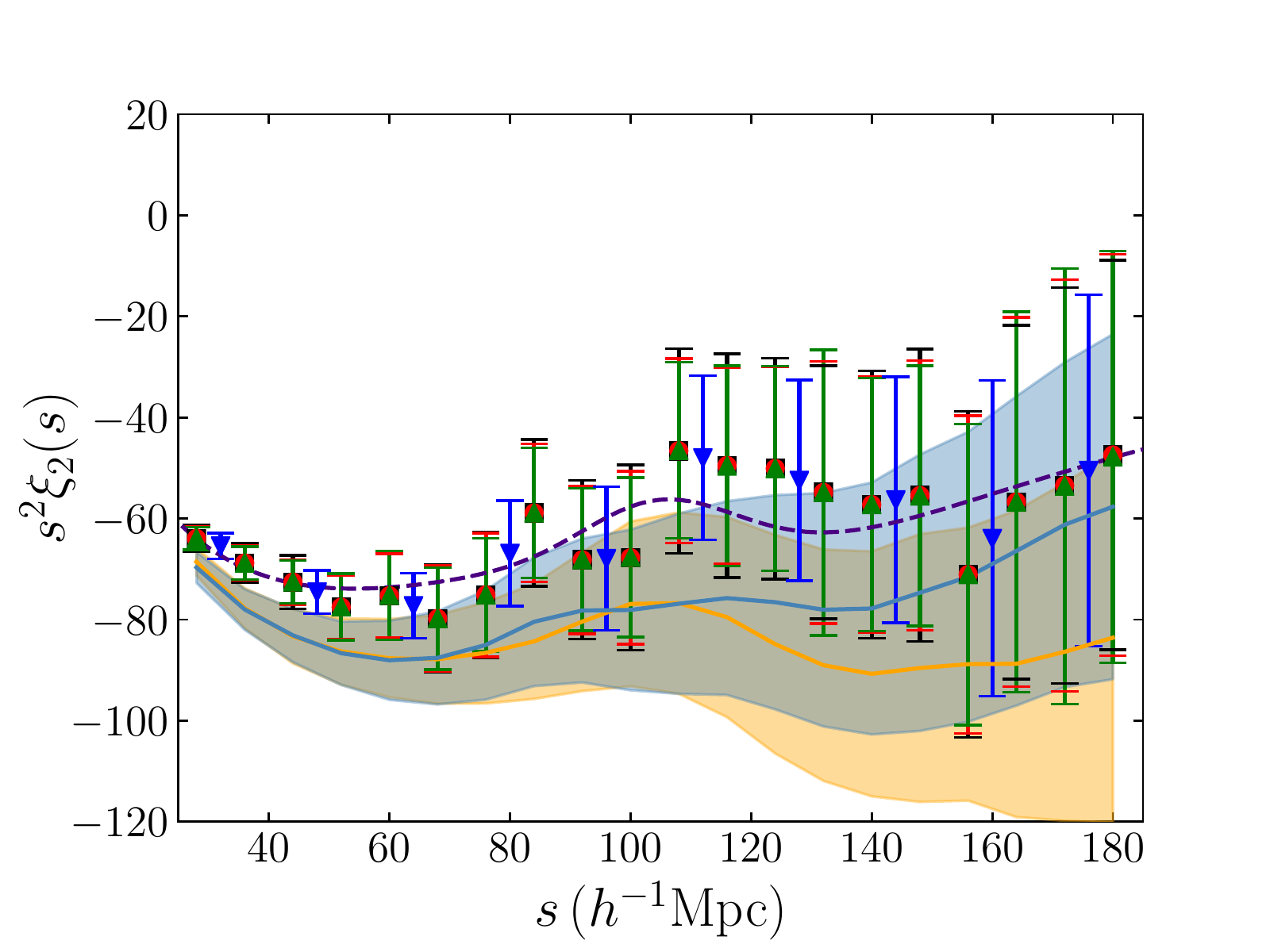}\vspace{-0.3cm}
\caption{Monopole (top) and quadrupole (bottom) auto-correlation functions of the BOSS CMASS galaxies (markers) computed using two different binning schemes (20 and 10 linear bins in $s$) coupled with the jackknife configurations given in Table\,\ref{tab:jkconfig} for the error estimation (200, 100, 50, 20 resamplings). We overplot in turquoise (orange) the mean $\pm\sigma$ values from the 1000 Patchy (LN) mocks. The analytic best-fit models to the CMASS measurements that we use to estimate the $\alpha$ shift parameter (see Sec.\,\ref{sec:analyticmodel}) are shown as dashed purple curves.}

\label{fig:2PCF}
\end{center}
\end{figure}

In Fig.\,\ref{fig:2PCF} we present the BOSS CMASS monopole and quadrupole two-point auto-correlation functions compared to the mean predictions from the 1000 Patchy and LN mocks, and the analytic model defined in \S\,\ref{sec:analyticmodel}. The CMASS error bars are inferred from the jackknife covariances based on the four configurations shown in Tab.\,\ref{tab:jkconfig} coupled with two different binning schemes (see Sec.\,\ref{sec:2pt}). For the Patchy and LN mocks we show the dispersion obtained from their 1000 realisations without applying any jackknife resampling. The LN mocks reproduce the BAO peak well, while the Patchy result differ from the CMASS measurements in the broadband shape. The LN monopole prediction tends to overestimate the observed clustering amplitude at $s\lesssim60\,h^{-1}$Mpc and beyond $150\,h^{-1}$Mpc. Patchy, instead, reproduces well the CMASS clustering up to $\sim80\,h^{-1}$Mpc and underestimates it beyond BAO scales. The systematic difference in shape will be accounted for by the nuisance parameters in the model and so it will not influence the analysis of the $\alpha$ shift parameter. We also overplot the analytic 2PCF model used in our MCMC algorithm to extract the $\alpha$ BAO parameter (see Sec.\,\ref{sec:javiermethod}). 
The  best-fit analytic model is in good agreement with the CMASS multipole measurements on all scales.

\begin{figure}
\begin{center}\vspace{-0.3cm}
\includegraphics[width=1.1\linewidth]{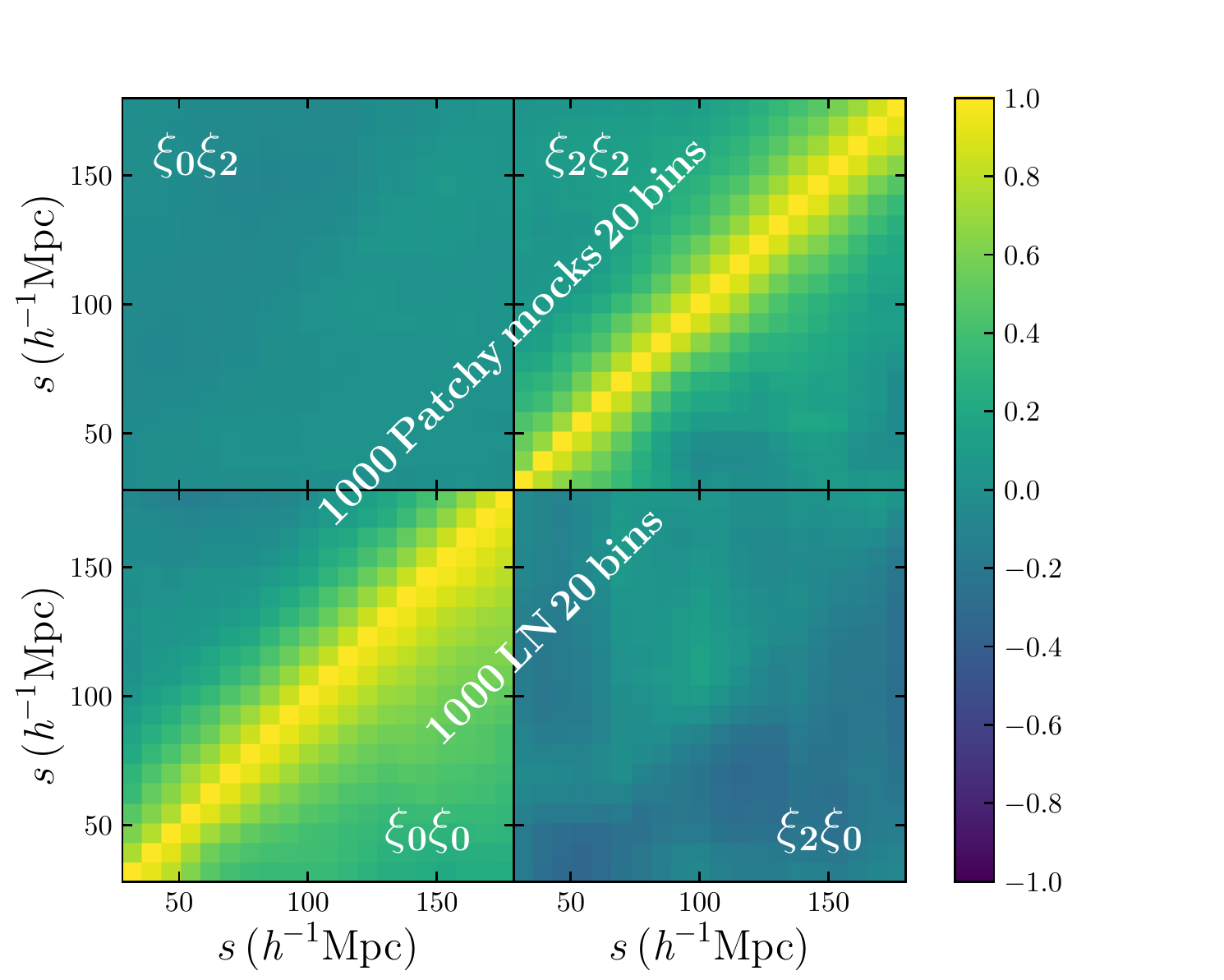}\vspace{-0.4cm}
\caption{Normalised monopole and quadrupole auto- and cross-covariances obtained from the 1000 Patchy (upper triangle) versus the 1000 LN (lower triangle) mocks, without applying any jackknife resampling. The normalisation is computed as $C_{\rm{ij}}^{\rm{norm}}=C_{\rm{ij}}/\sqrt{C_{\rm{ii}}\,C_{\rm{jj}}}$, where $C_{\rm{ij}}$ is given in Eq.\,\ref{eq:covmocks}. The mean value and 1\,$\sigma$ dispersion of these mocks are shown in Fig.\,\ref{fig:2PCF} as solid lines with the corresponding shaded region.}
\label{fig:matrixmock}
\end{center}
\end{figure}
\begin{figure*}
\begin{center}\vspace{-0.4cm}
\includegraphics[width=0.41\linewidth]{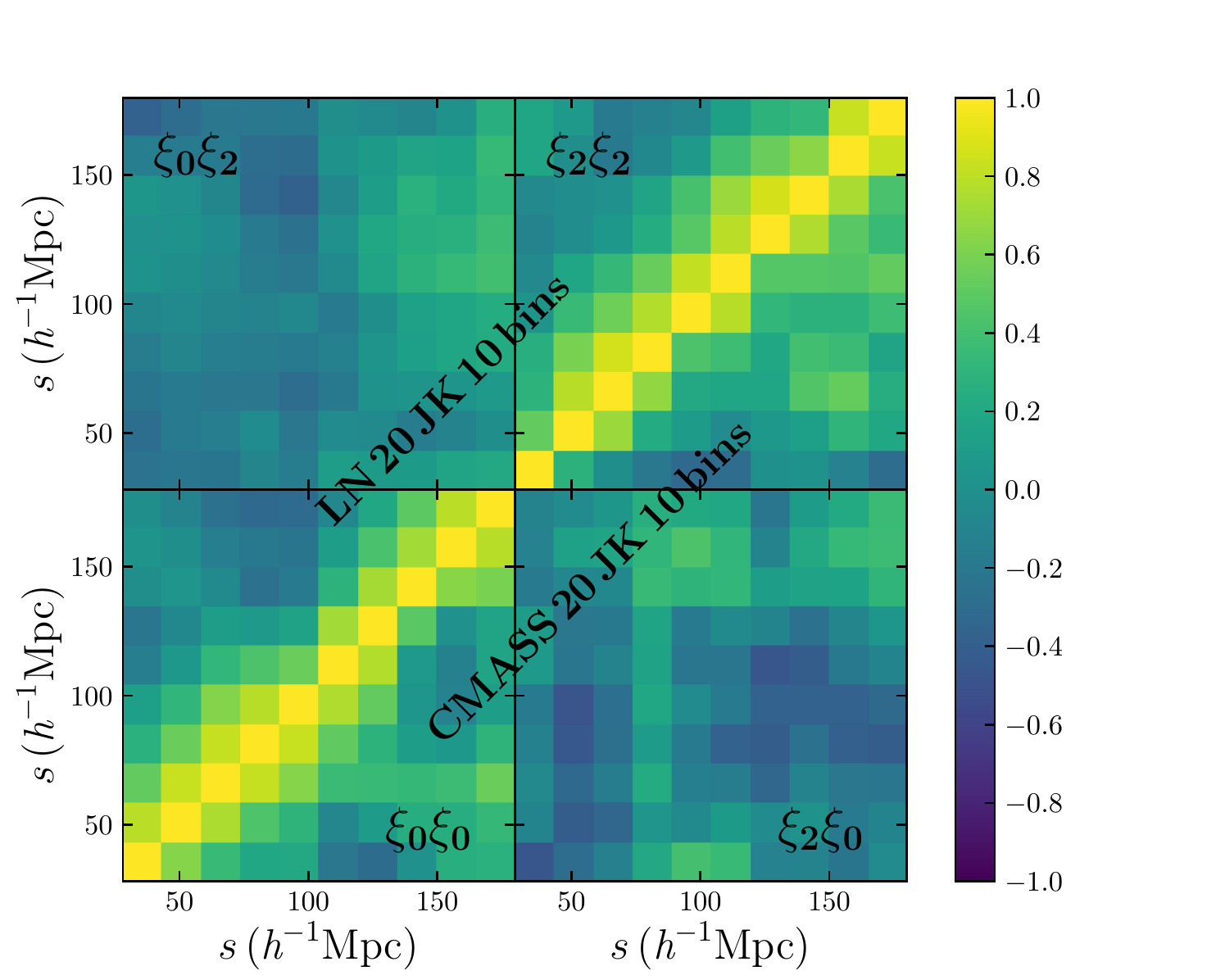}\quad
\includegraphics[width=0.41\linewidth]{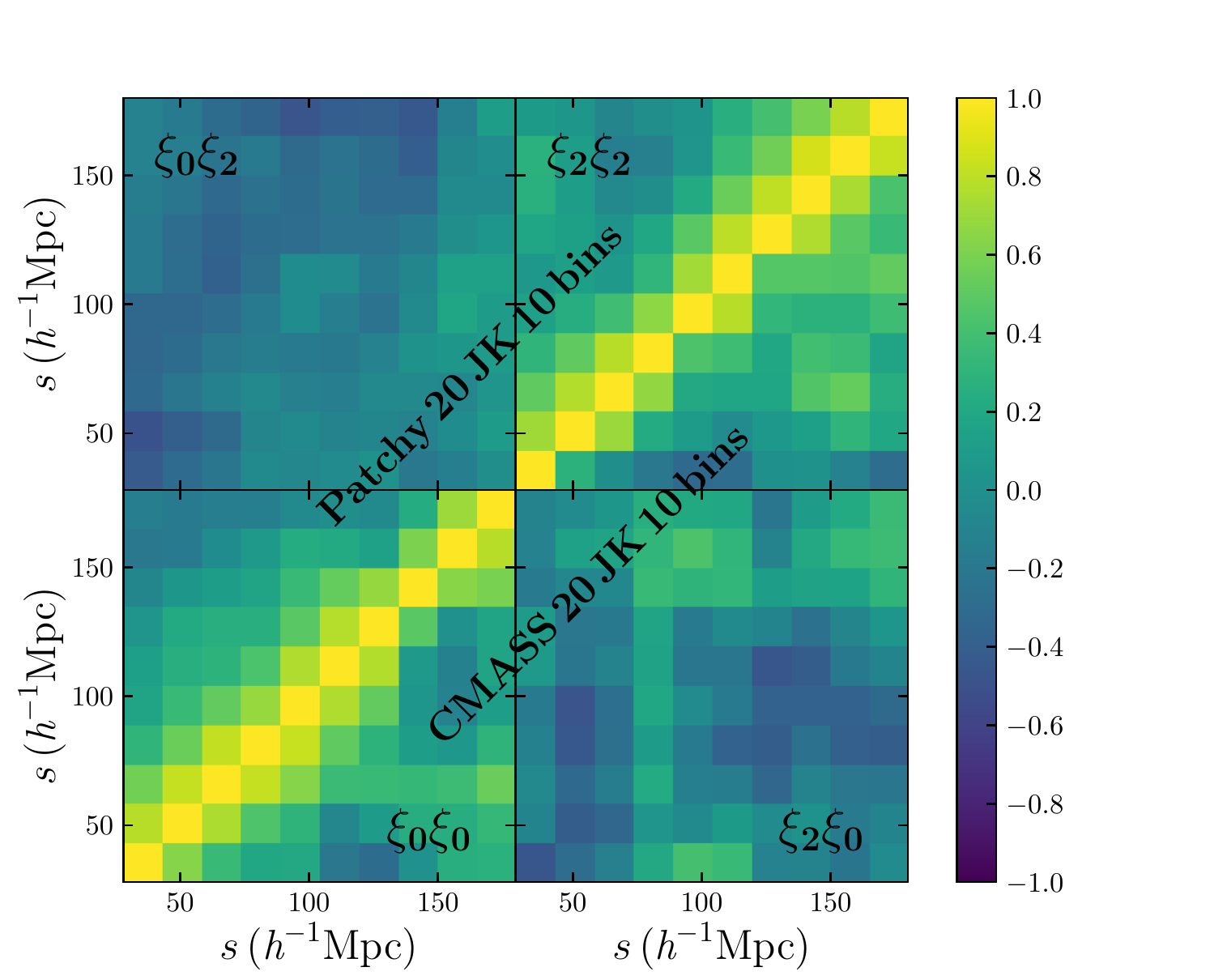}\vspace{-0.1cm}
\includegraphics[width=0.41\linewidth]{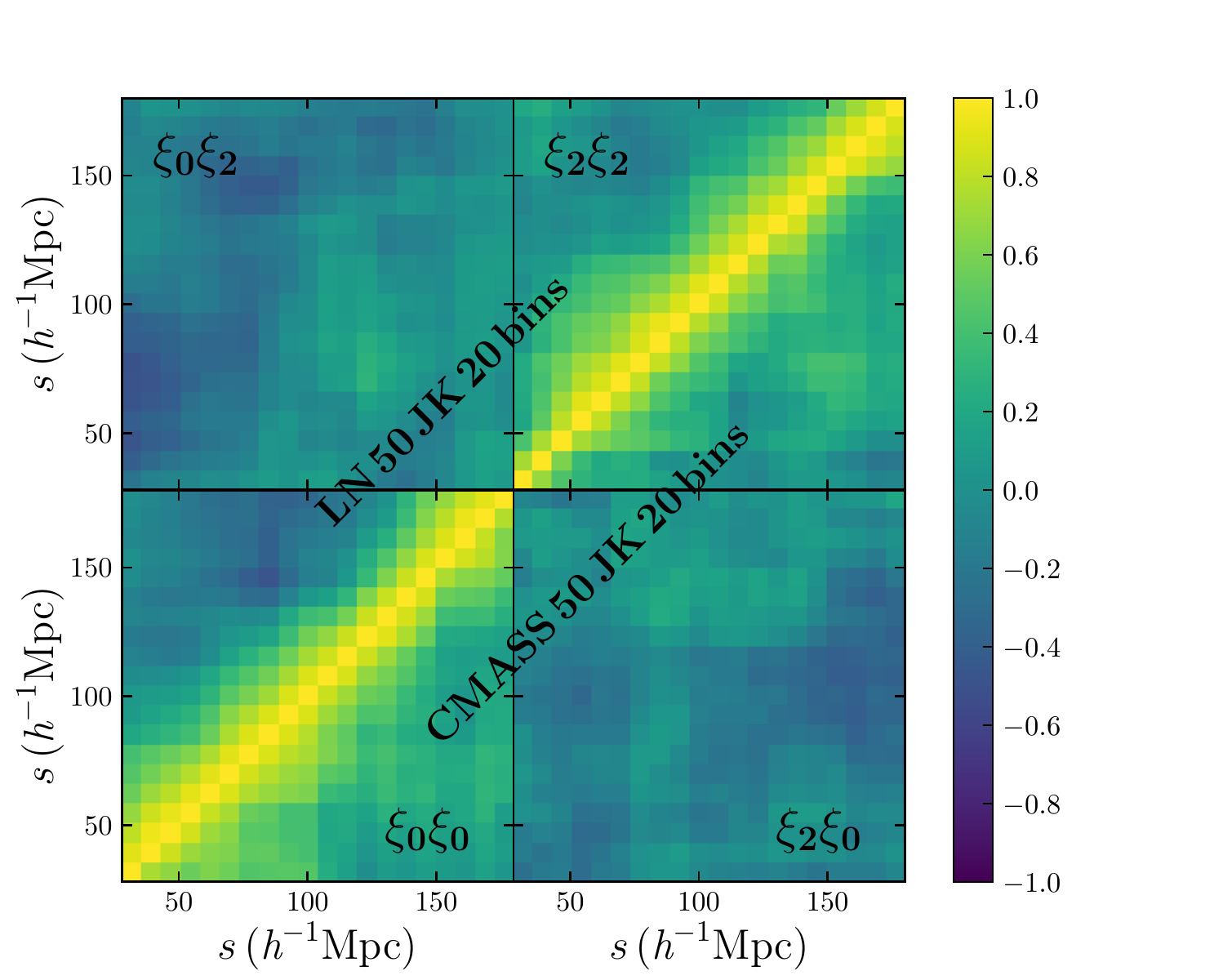}\quad
\includegraphics[width=0.41\linewidth]{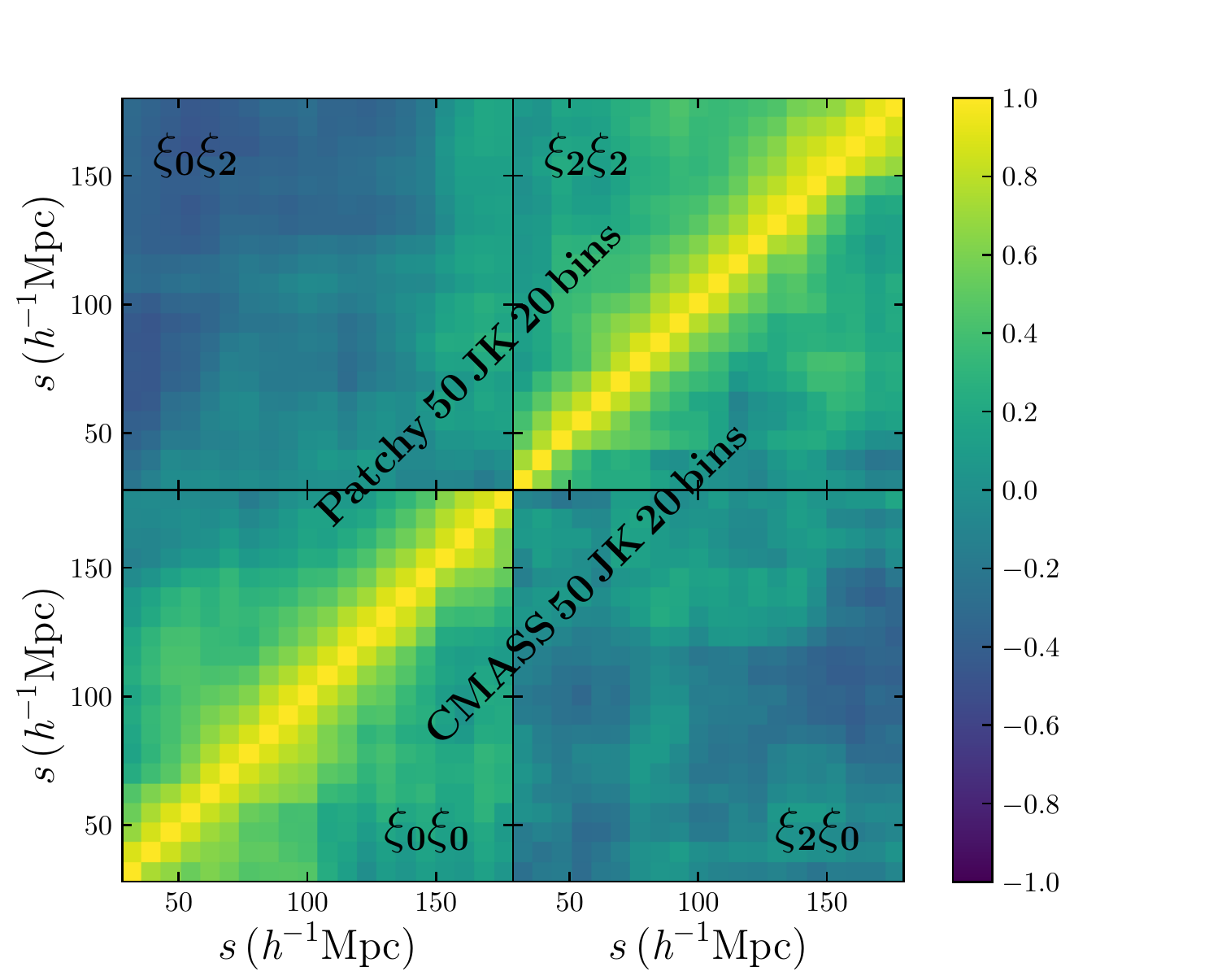}\vspace{-0.1cm}
\includegraphics[width=0.41\linewidth]{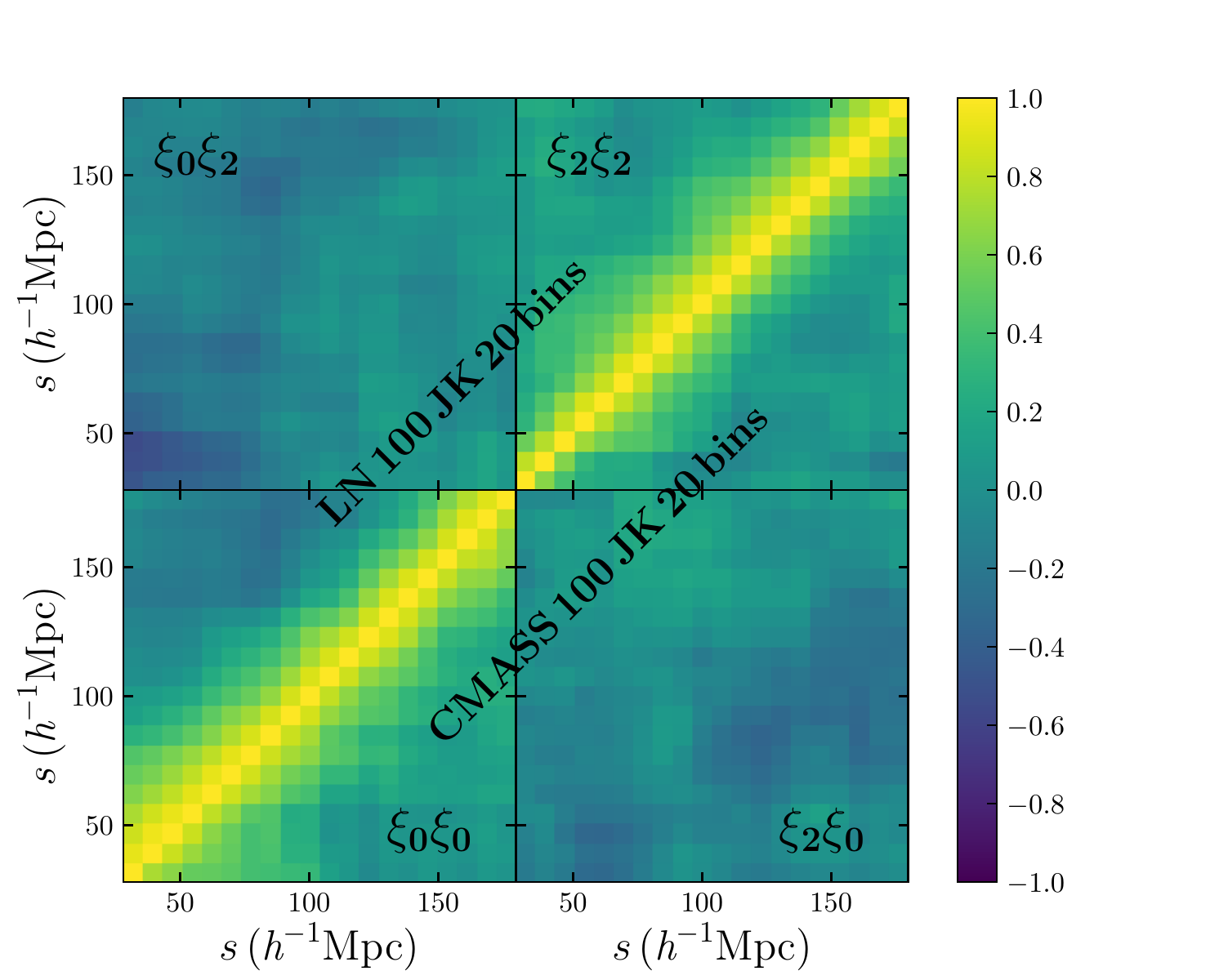}\quad
\includegraphics[width=0.41\linewidth]{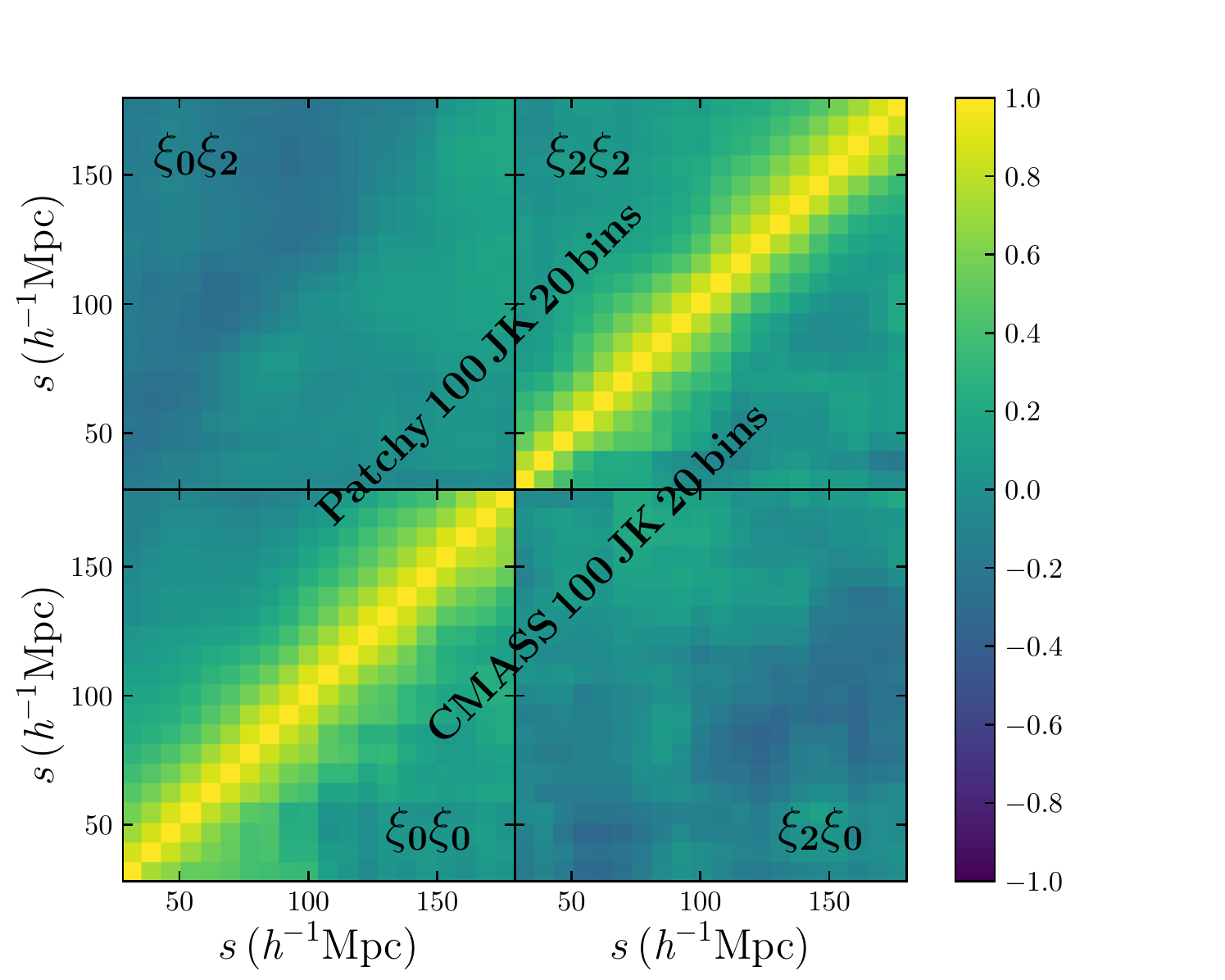}\vspace{-0.1cm}
\includegraphics[width=0.41\linewidth]{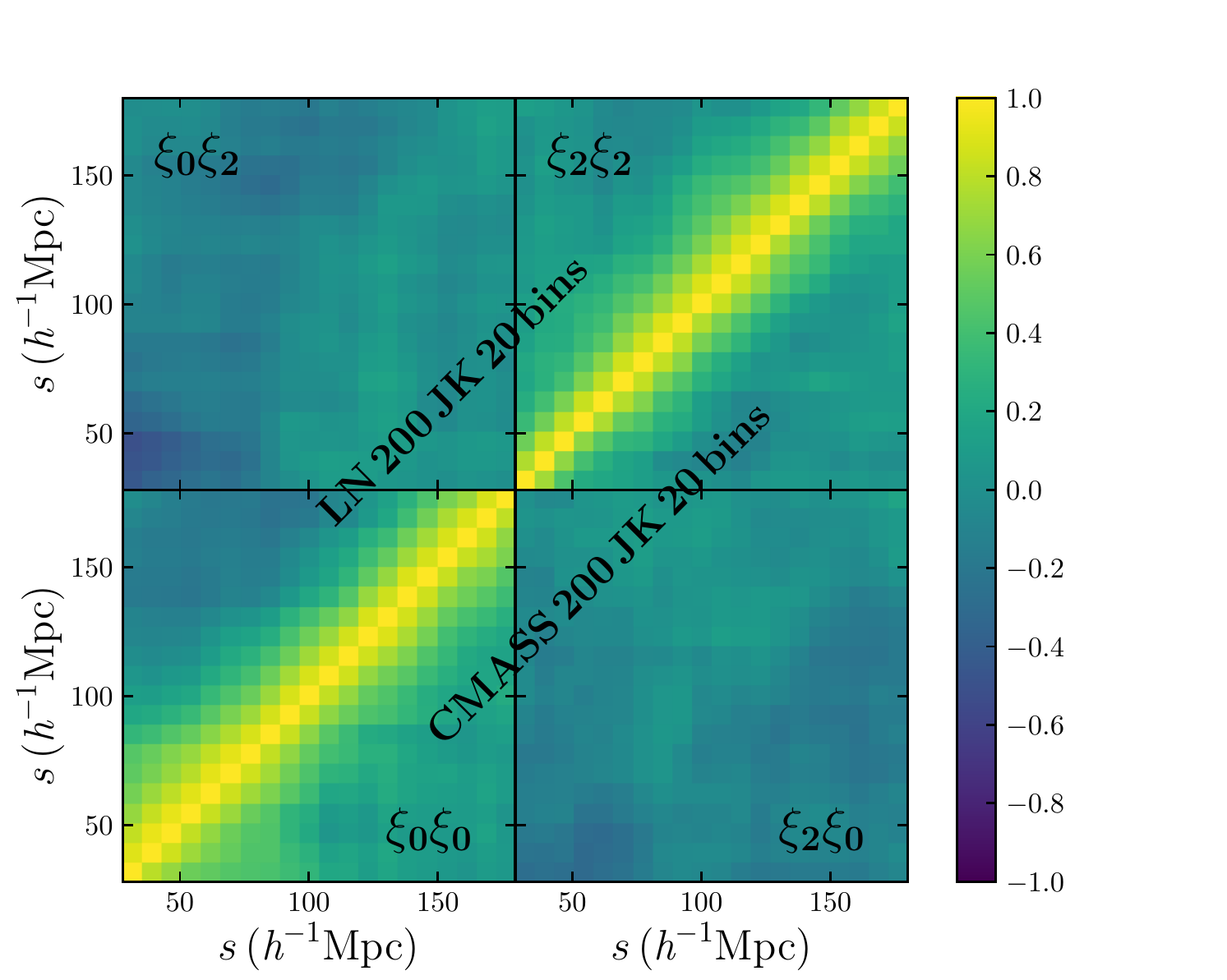}\quad
\includegraphics[width=0.41\linewidth]{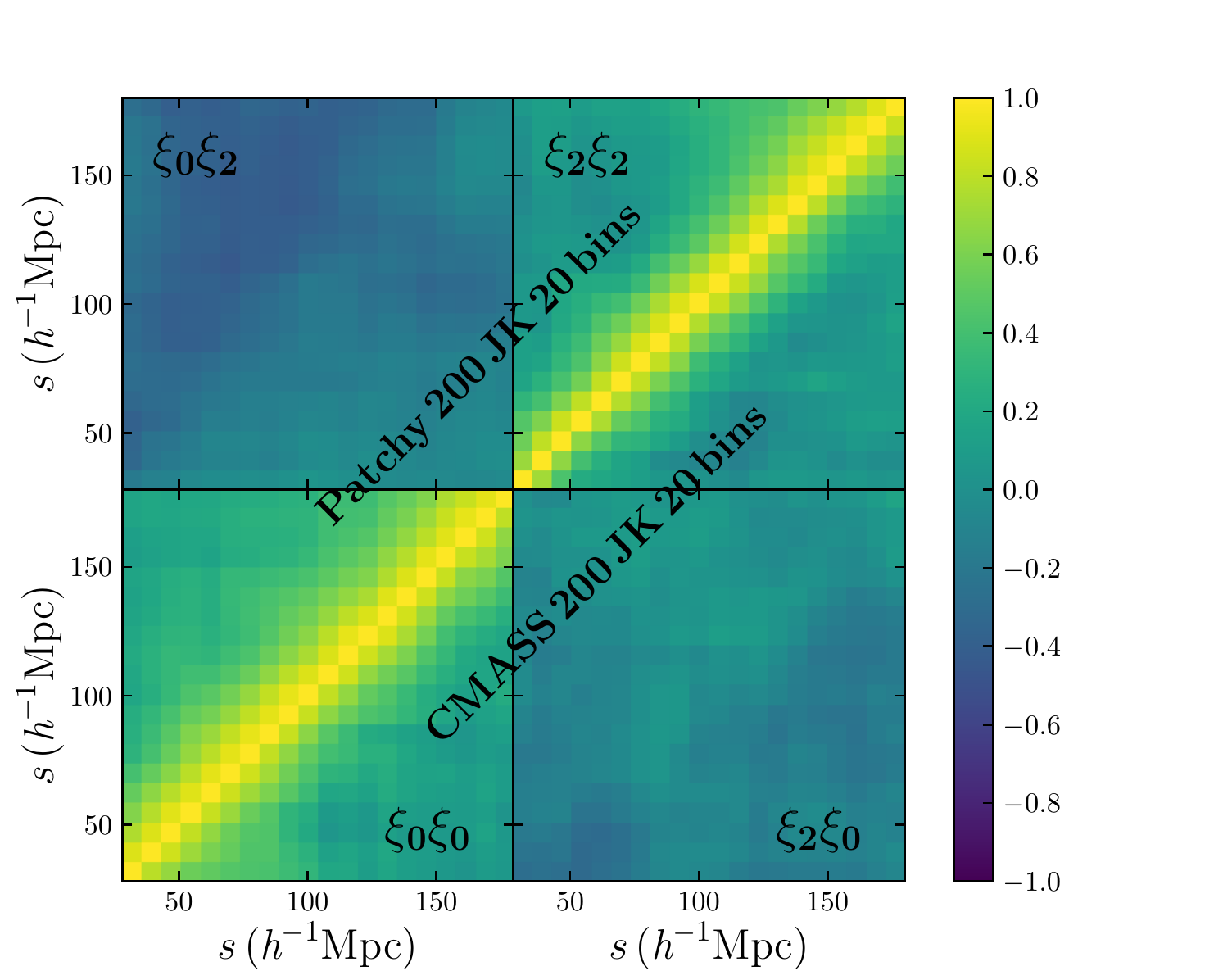}
\caption{Normalised covariances obtained from jackknife resampling performed on 10 different LN and Patchy realisations (upper triangles) and on BOSS CMASS data (lower triangles). For both the LN and Patchy results we show the average estimate from the 10 covariances. From top to bottom we display the 20, 50, 100 and 200 jackknife configurations coupled with two binning schemes.}
\label{fig:matricesJK}
\end{center}
\end{figure*}

In Fig.\,\ref{fig:matrixmock} we compare the normalised monopole and quadrupole covariances from the 1000 Patchy (upper triangle) and the 1000 LN (lower triangle) mocks. These measurements are both built in 20 $s$ bins, without applying any jackknife resampling. The matrix is normalised as $C_{\rm{ij}}^{\rm{norm}}=C_{\rm{ij}}/\sqrt{C_{\rm{ii}}\,C_{\rm{jj}}}$, with $C_{\rm{ij}}$ given in Eq.\,\ref{eq:covmocks}. Despite the number of Patchy and LN mocks used here is the same, the LN result shows a higher covariance in the off-diagonal terms. The LN mocks are based on a number of approximations which prevent them from capturing the full non-linear effects, which instead are included in the Patchy mocks \citep[see also][]{Blot2019}.

\begin{figure*}
\begin{center}\vspace{-0.4cm}
\includegraphics[width=0.41\linewidth]{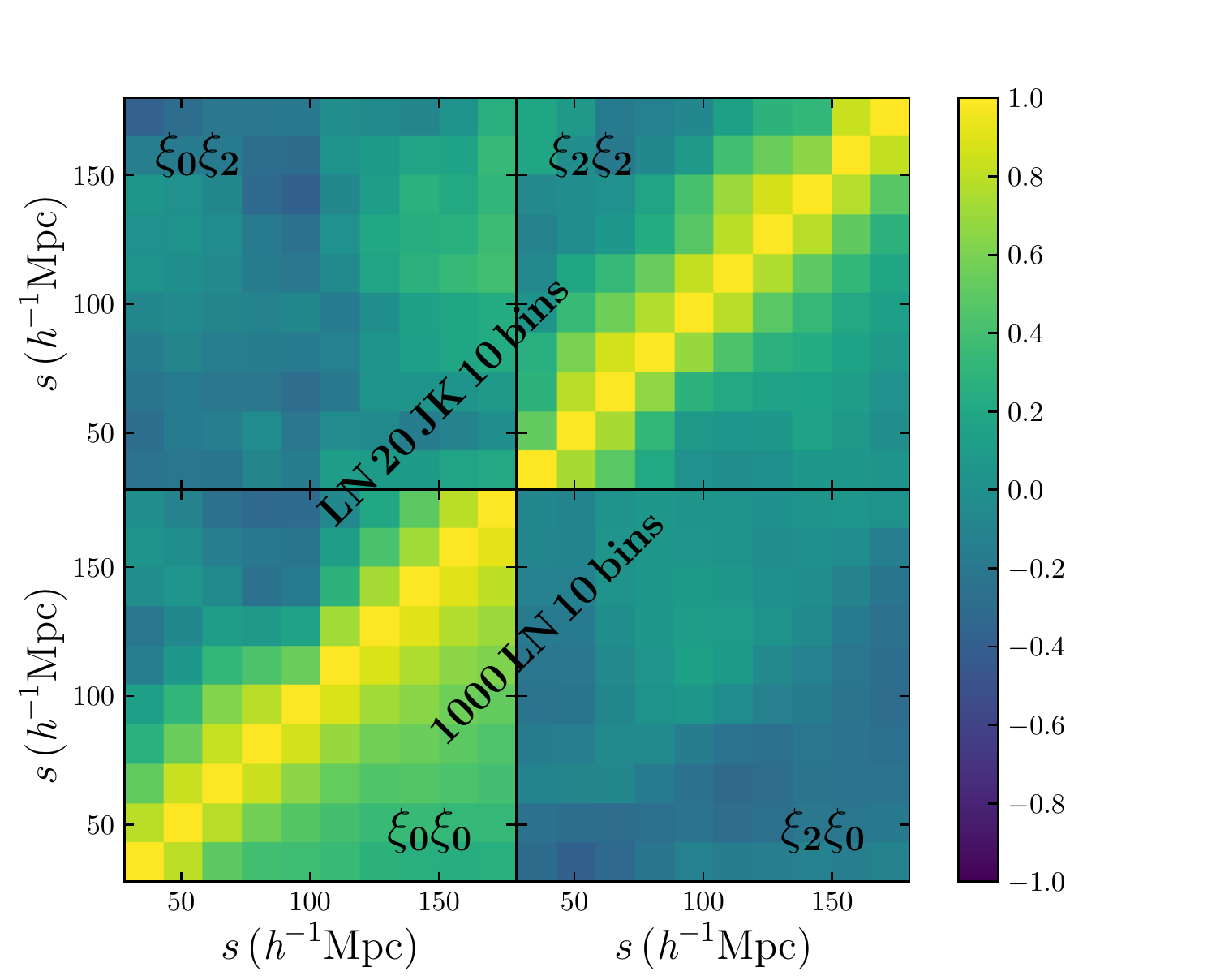}\quad
\includegraphics[width=0.41\linewidth]{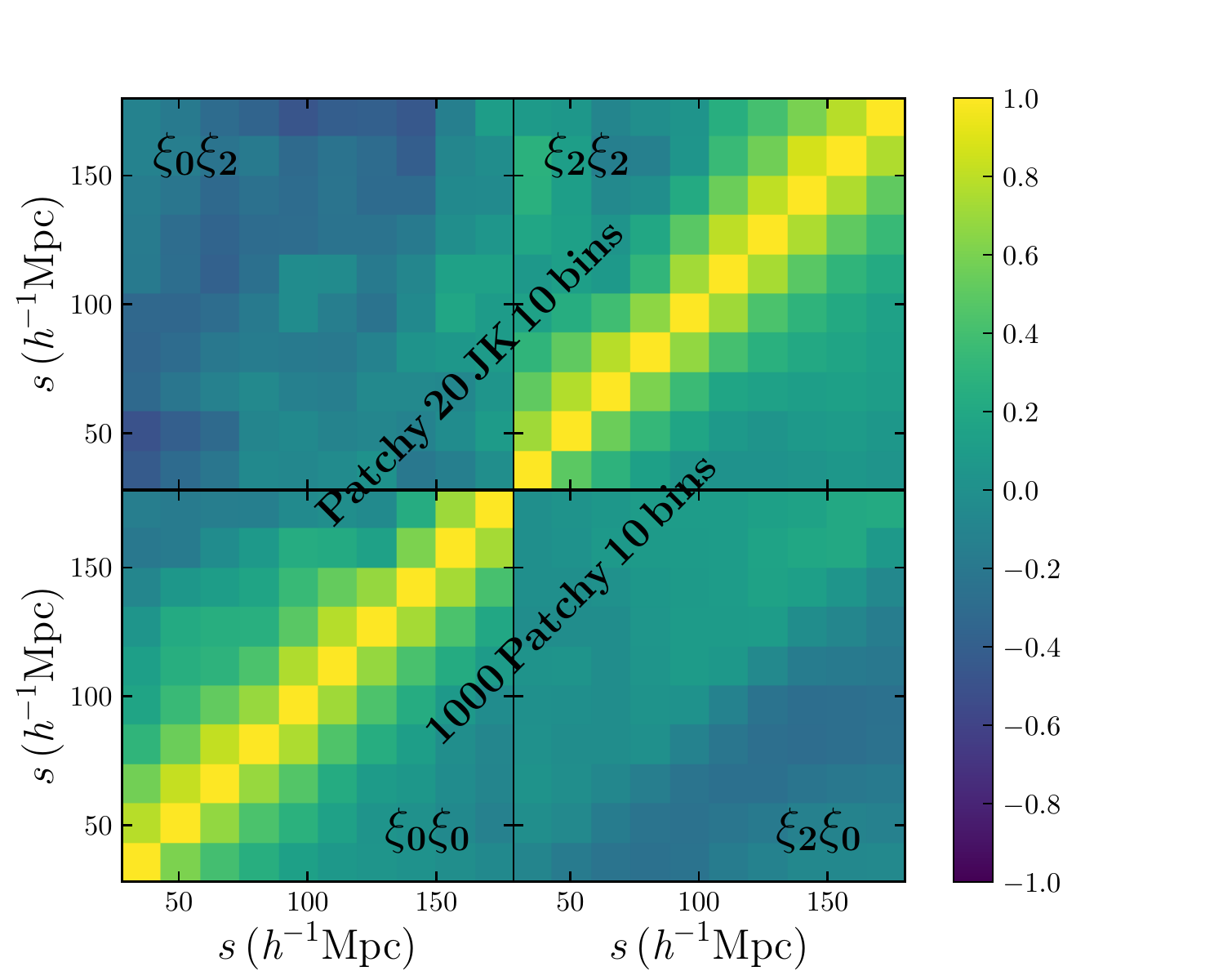}\vspace{-0.1cm}
\includegraphics[width=0.41\linewidth]{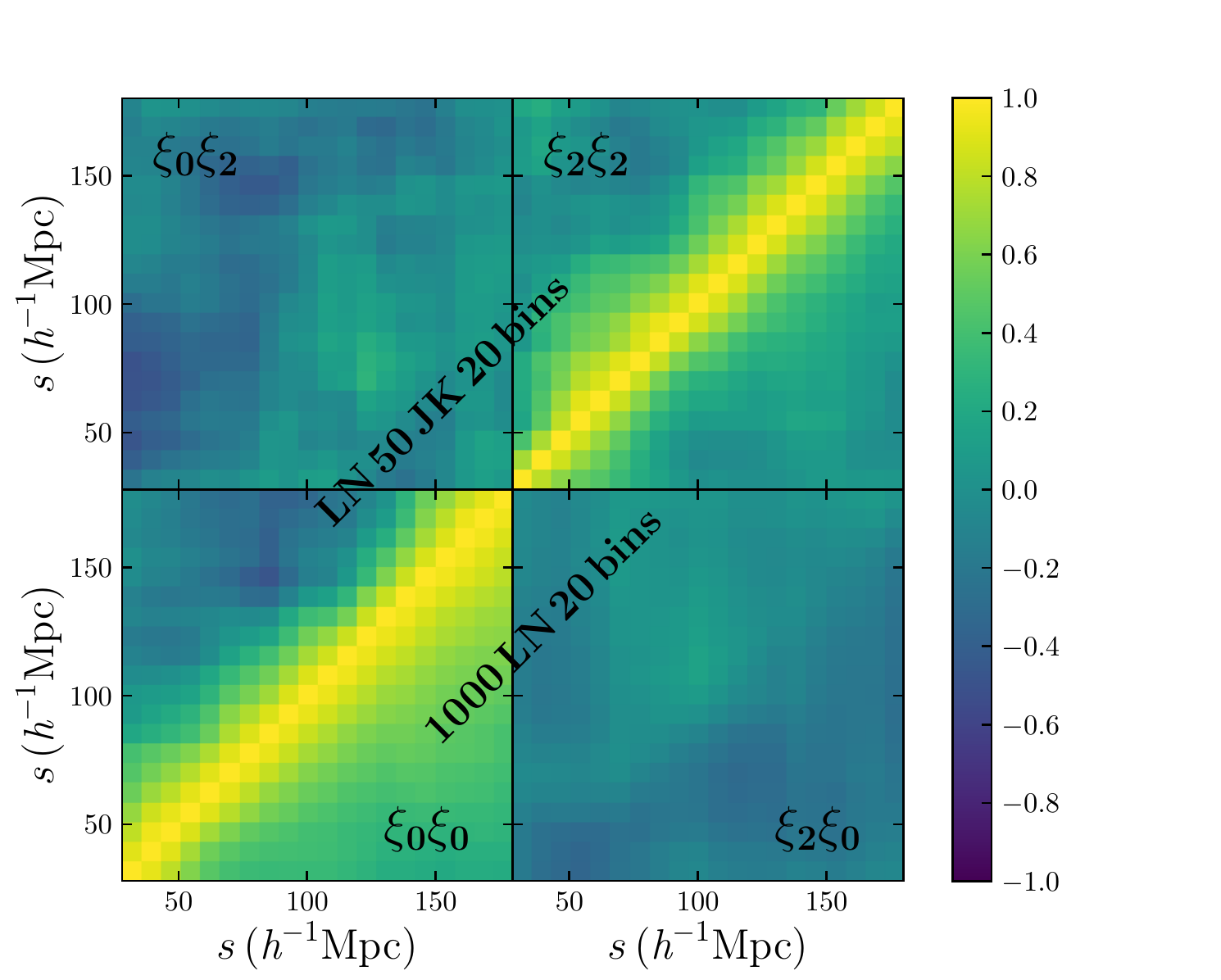}\quad
\includegraphics[width=0.41\linewidth]{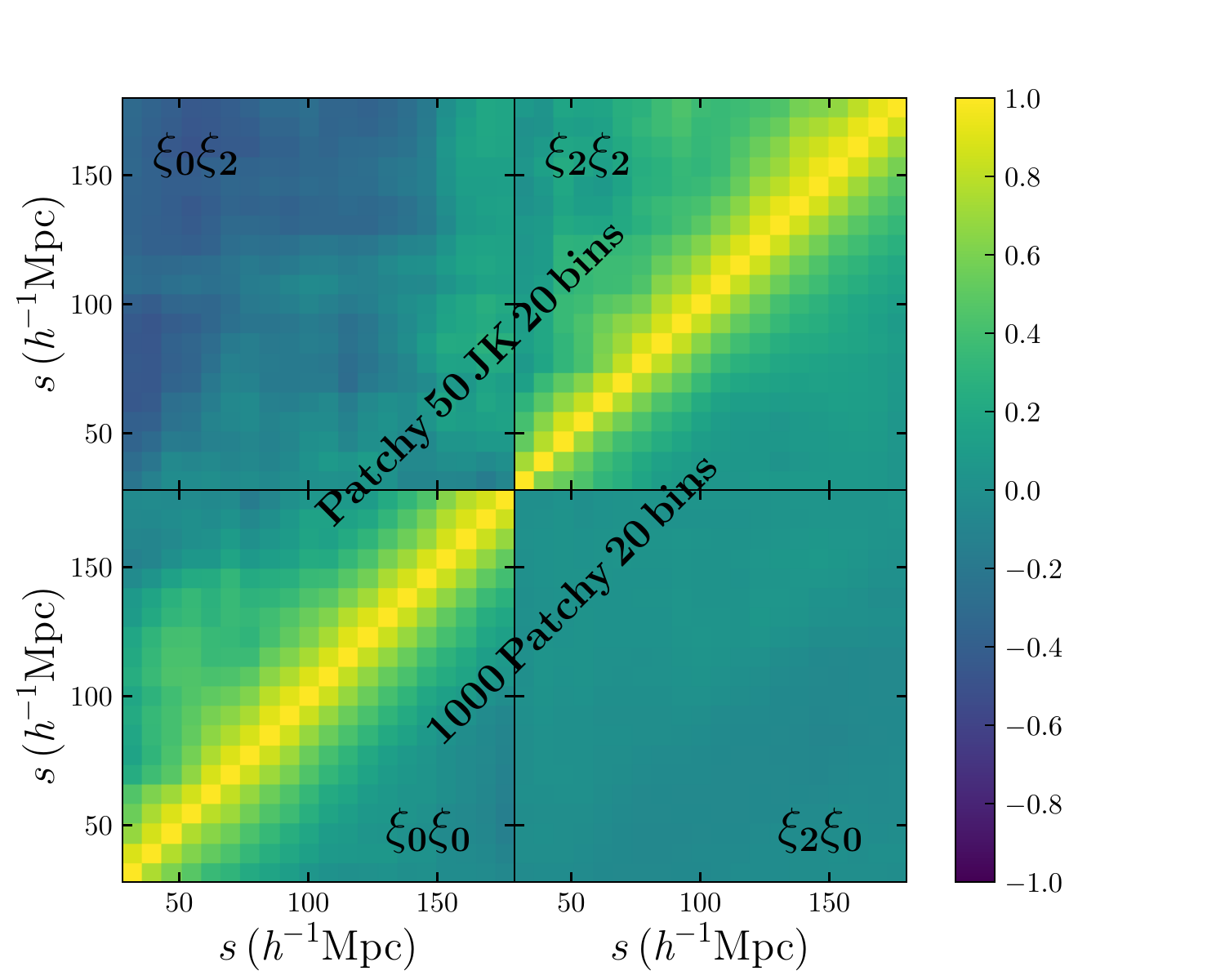}\vspace{-0.1cm}
\includegraphics[width=0.41\linewidth]{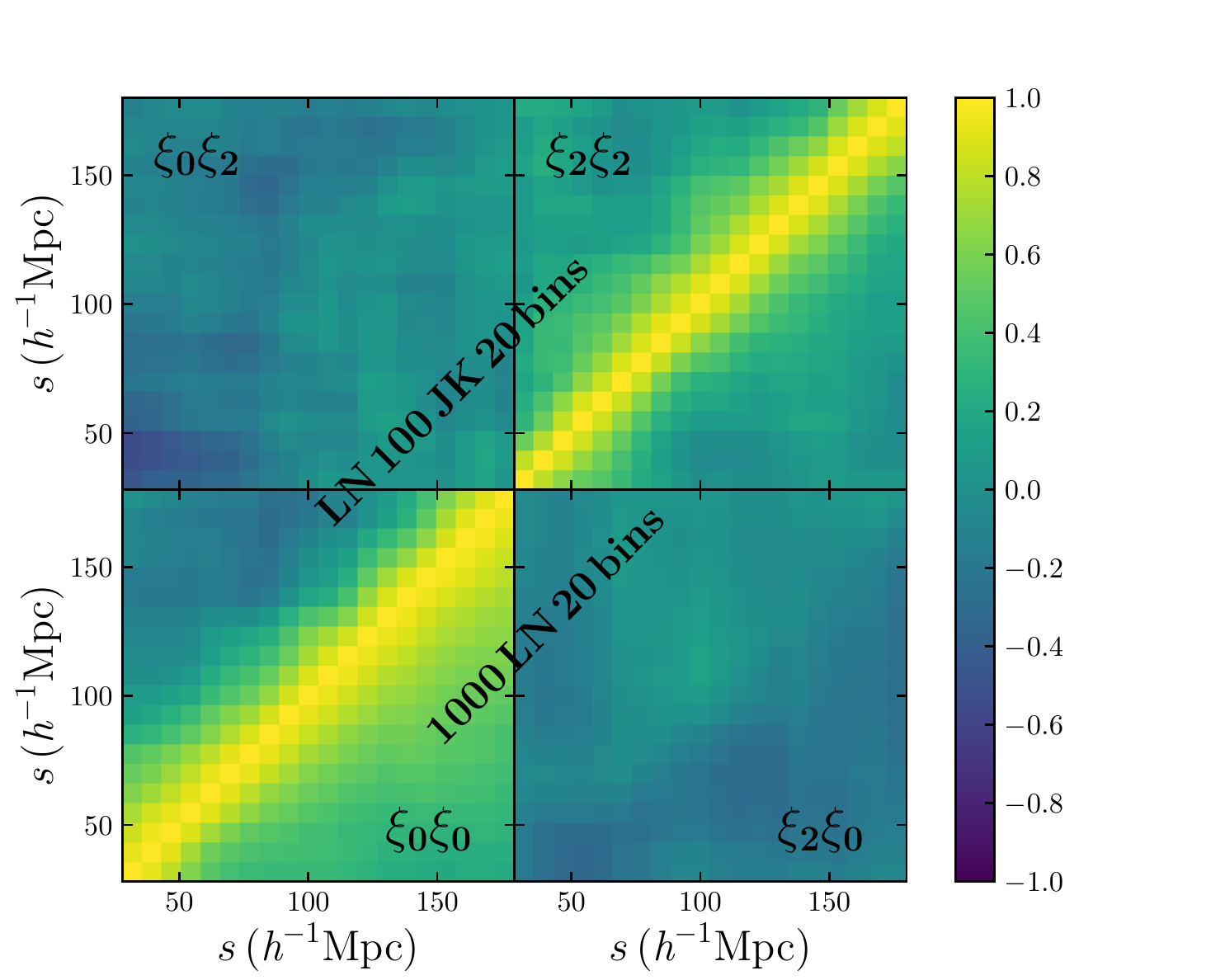}\quad
\includegraphics[width=0.41\linewidth]{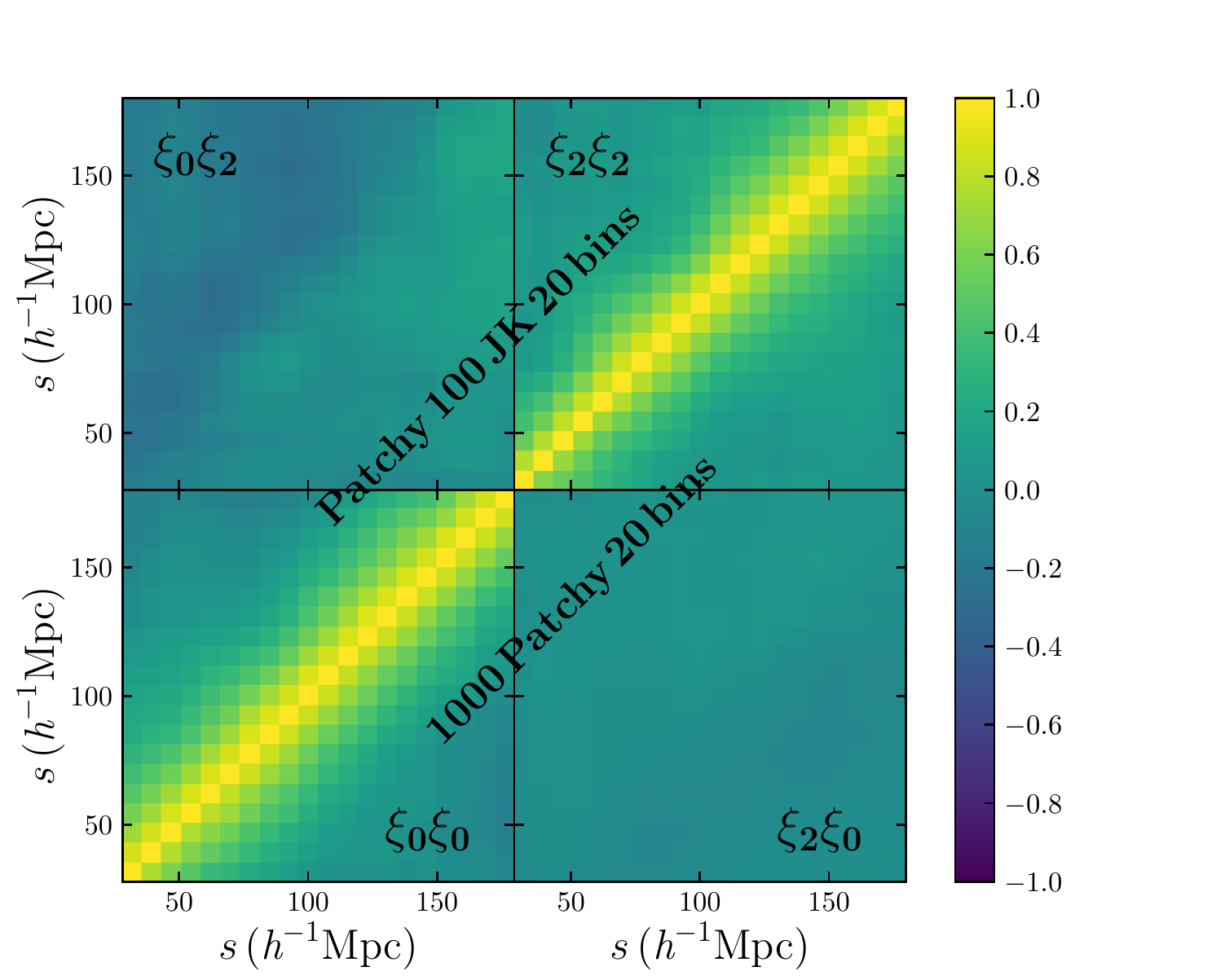}\vspace{-0.1cm}
\includegraphics[width=0.41\linewidth]{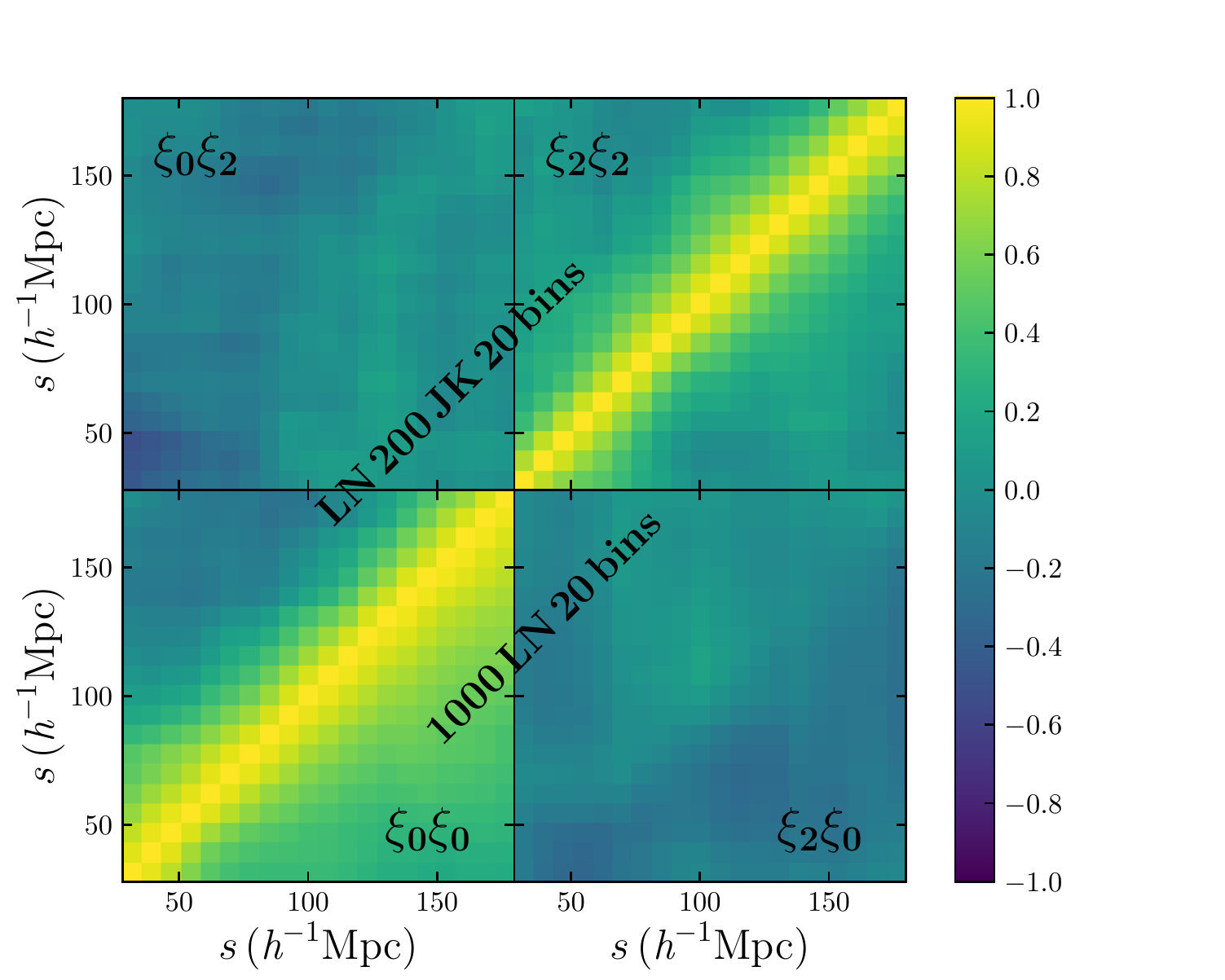}\quad
\includegraphics[width=0.41\linewidth]{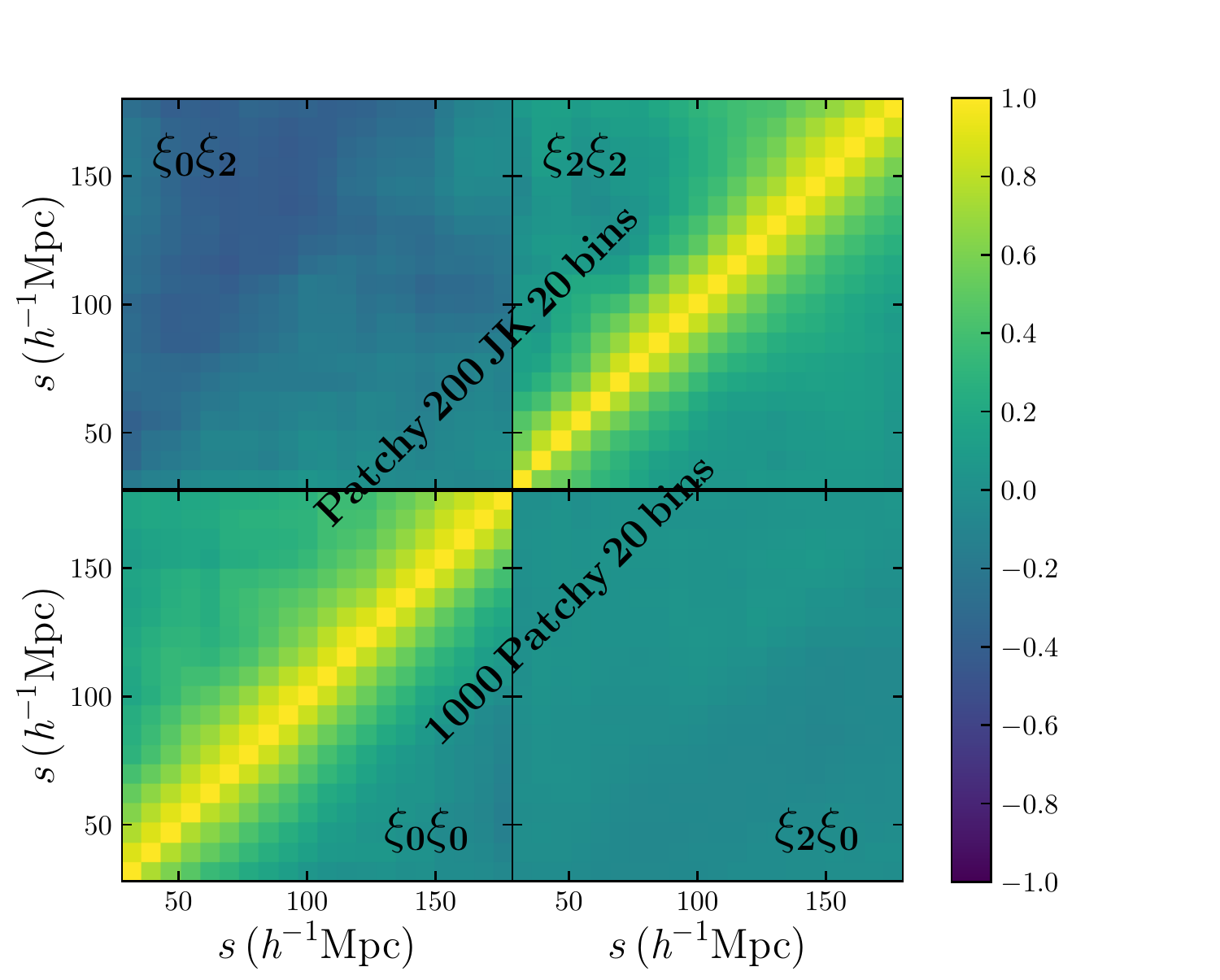}
\caption{Same result as Fig.\,\ref{fig:matricesJK}, but here in the lower triangles we show the reference covariances from 1000 LN and Patchy mocks without jackknife, which are also shown in Fig.\,\ref{fig:matrixmock}. Normalised covariances obtained from jackknife resampling performed on 10 different LN and Patchy realisations (upper triangles) and on BOSS CMASS data (lower triangles). In the upper triangles we show the average estimate from the covariances of 10 mock realisations.}
\label{fig:matricesJKvsideal}
\end{center}
\end{figure*}

In the left column of Fig.\,\ref{fig:matricesJK}, we compare the average normalised covariances obtained by performing jackknife resamplings on 10 different LN realisations (upper triangles) versus BOSS CMASS data (lower triangles). In the right column are the average covariances of 10 different Patchy mocks (upper triangles) and CMASS data (lower). In each column, from top to bottom, we show the 20, 50, 100 and 200 jackknife configurations listed in Tab.\,\ref{tab:jkconfig}, respectively coupled with 10, 20, 20, 20 linear bins in $s$ (see Sec.\,\ref{sec:measurements}). The normalisation is calculated as described in Fig.\,\ref{fig:matrixmock}, with $C_{\rm{ij}}$ given by Eq.\,\ref{eq:covma}. 
It is evident that the noise in the covariance estimates is reduced as the number of resamplings is increased.
The covariances in 20 bins obtained from 200 jackknife resamplings on the Patchy, LN mocks or CMASS observations are consistent with the results from 1000 Patchy and LN mocks without jackknife shown in Fig.\,\ref{fig:matrixmock}. These covariances lead to consistent error bars on the galaxy clustering multipoles, as we can appreciate in Fig.\,\ref{fig:2PCF}.

In Fig.\,\ref{fig:matricesJKvsideal} we compare the average normalised jackknife covariances from LN and Patchy mocks versus their reference covariances from 1000 independent realisations already shown in Fig.\,\ref{fig:matrixmock}. There is a systematic difference in the covariances. The noise in the off-diagonal terms of the JK covariances is larger compared to the reference covariances and reduces as the number of jackknife regions increases.

\begin{figure*}
\begin{center}
\includegraphics[width=0.45\linewidth]{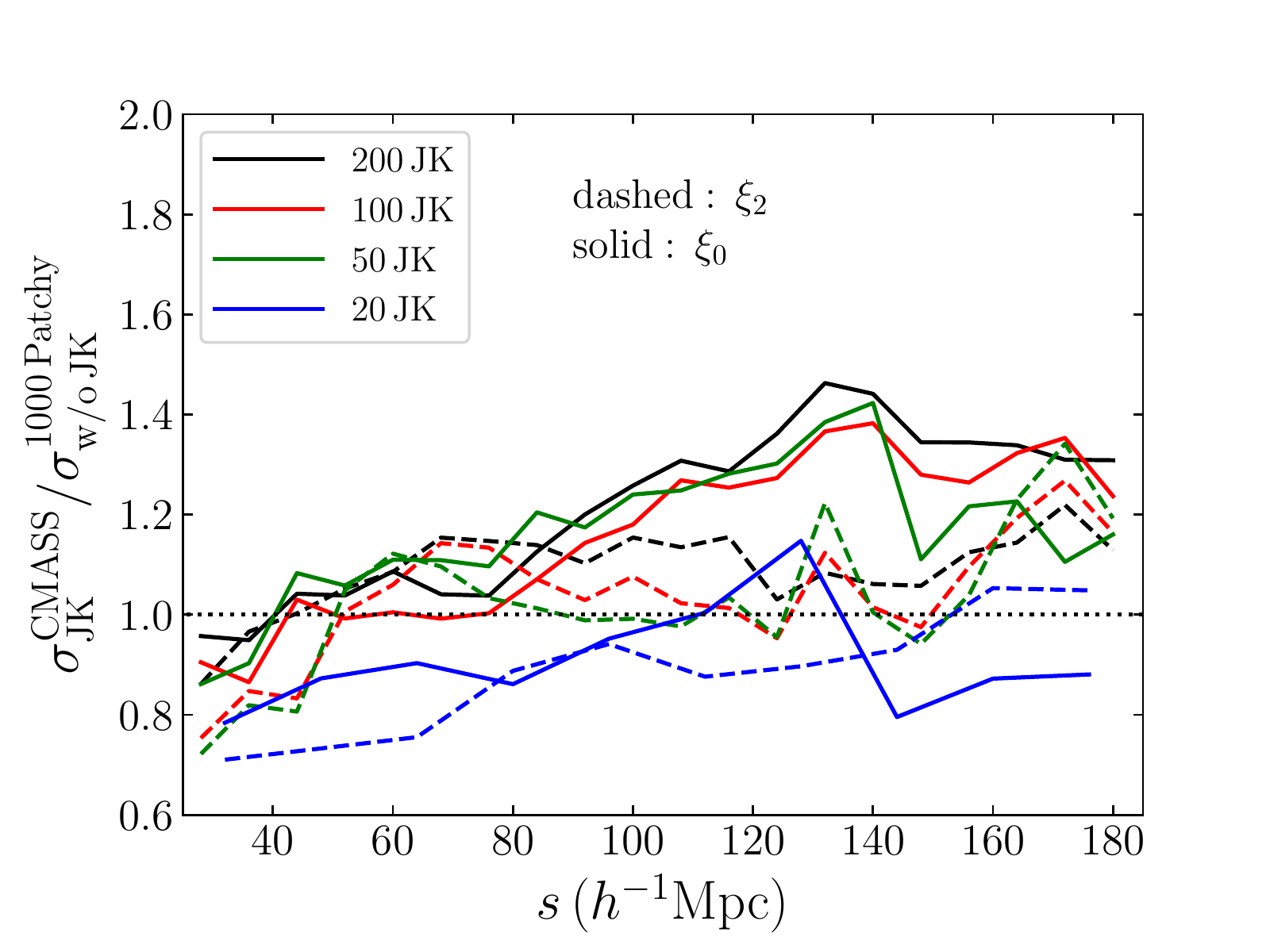}
\includegraphics[width=0.45\linewidth]{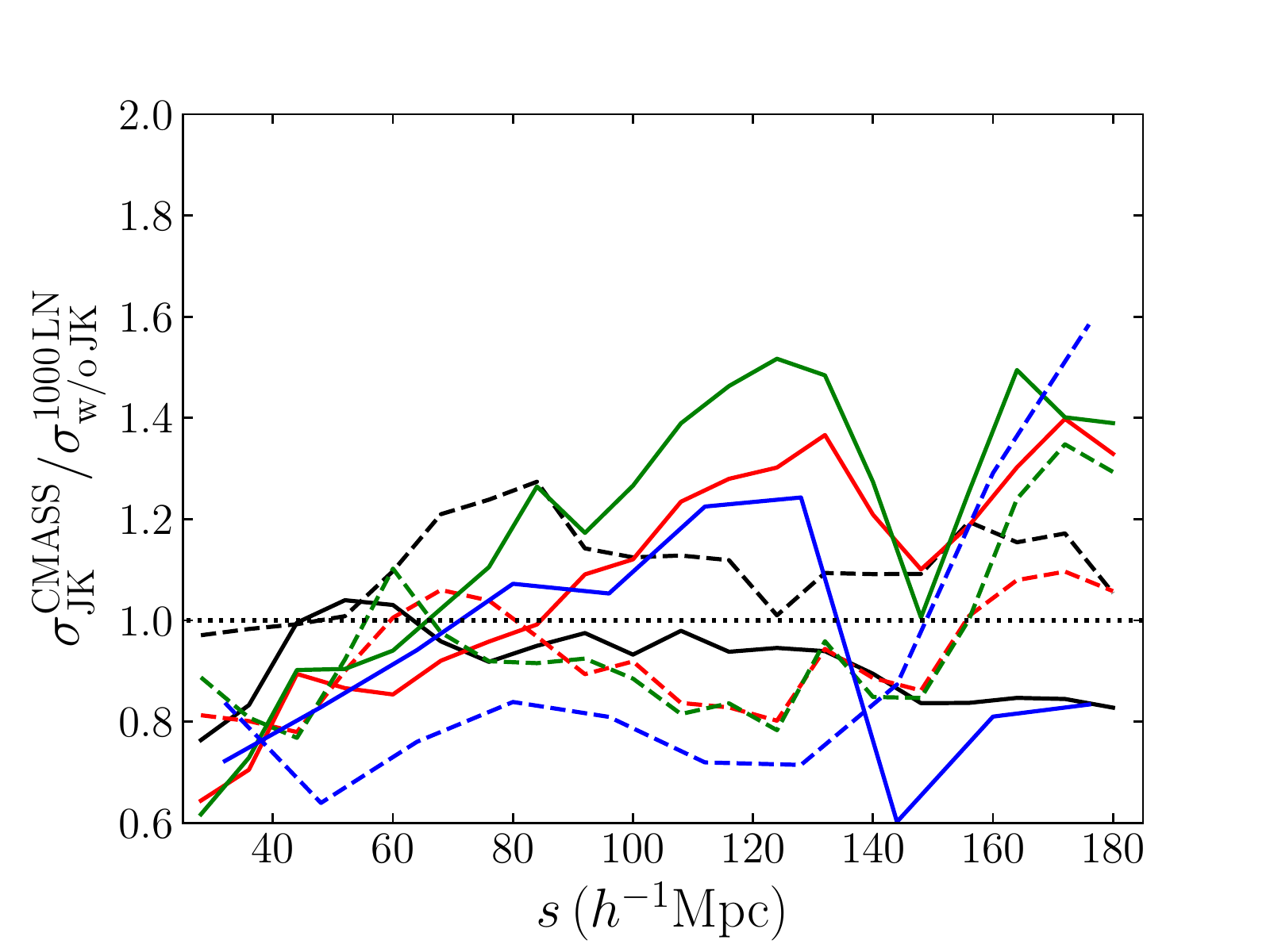}
\includegraphics[width=0.45\linewidth]{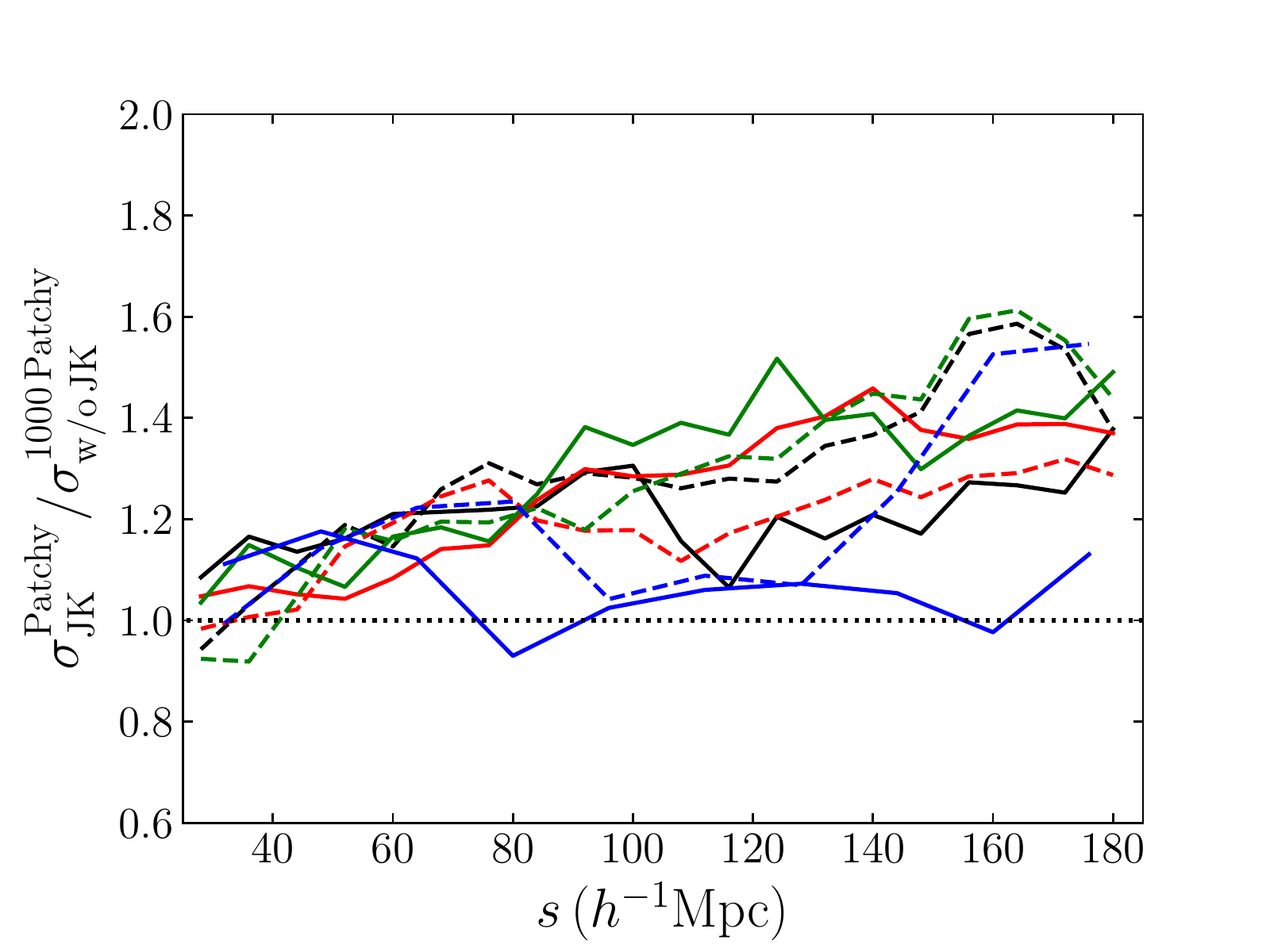}
\includegraphics[width=0.45\linewidth]{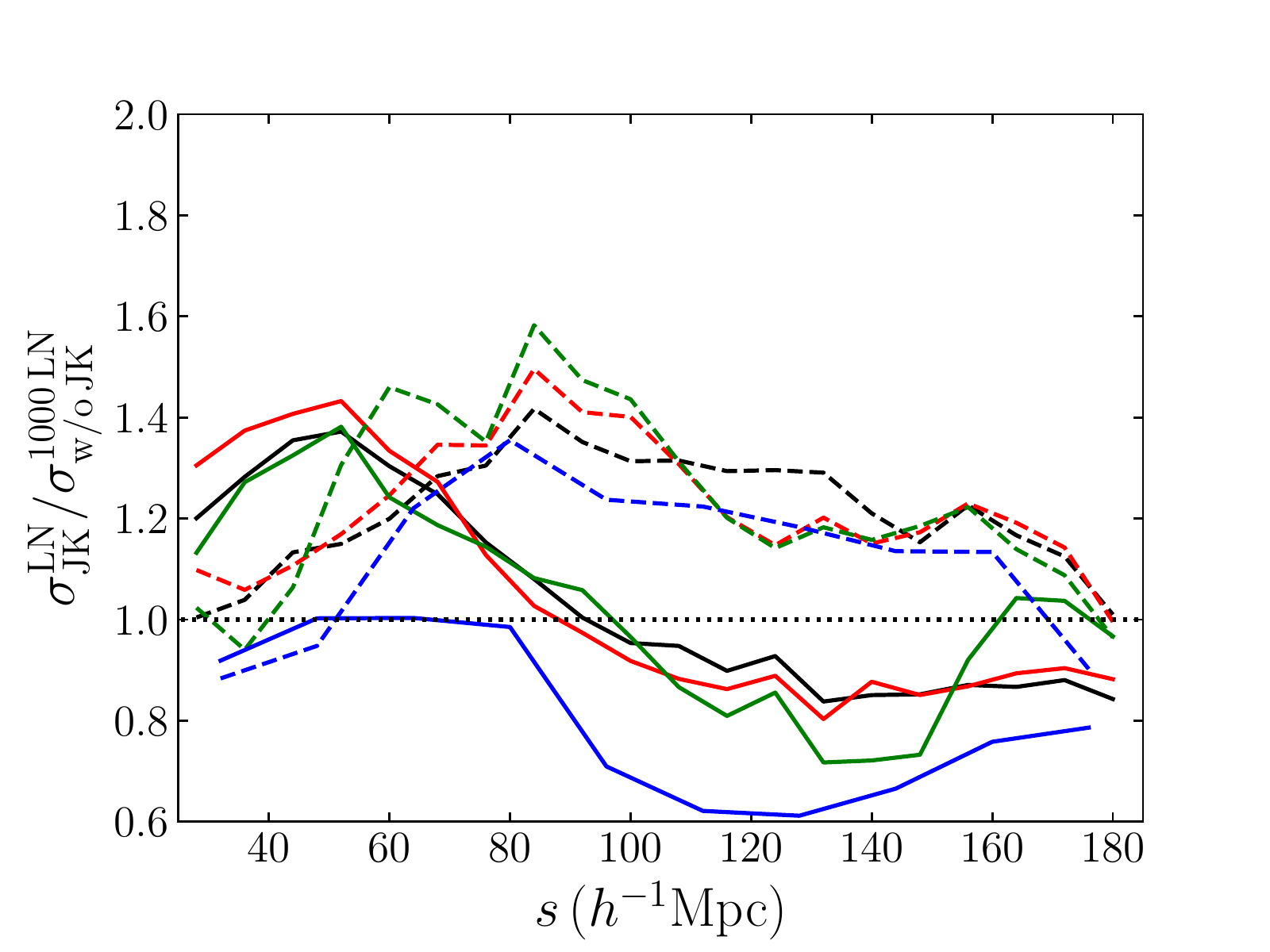}
\caption{\textit{Top row:} Ratios between the 1$\sigma$ uncertainties obtained from the CMASS jackknife covariances and the 1000 Patchy (left) or LN (right) mocks without JK. The solid (dashed) lines correspond to the monopole (quadrupole) measurements. \textit{Bottom row:} Ratios between the average 1\,$\sigma$ errors obtained from jackknife resamplings on 10 different Patchy (left) and the LN (right) realisations and those from 1000 Patchy or LN without jackknife. We remind the reader that the 200, 100 and 50\,JK configurations are coupled with 20 $s$ bins, while the 20\,JK case with 10 bins. The monopole and quadrupole results are shown as solid and dashed lines, respectively. The horizontal dotted lines are shown to help the comparison.}
\label{fig:sigma}
\end{center}
\end{figure*}

Fig.\,\ref{fig:sigma} shows, as a function of the scale, the ratios of the jackknife variances and the 1000 independent Patchy or LN mocks. In the top panels are the results from CMASS jackknife against the reference mocks. The bottom panels show the average uncertainties from jackknife on 10 Patchy/LN realisations against the 1000 reference mocks without jackknife.  

We remind the reader that the 200, 100 and 50 JK configurations are measured in 20 $s$ bins, while the 20 JK case in 10 bins. The combined action of the jackknife size, number and binning is what determines the level of noise in the covariances. In Fig.\,\ref{fig:sigma} we see that the 20\,JK scheme always leads to the largest fluctuations in the $\sigma$ estimate due to having relatively few jackknife resamplings available. However, doing only 10 bins instead of 20 helps to partially mitigate these fluctuations.

 On small scales, the errors from CMASS jackknife shown in the top panels of Fig.\,\ref{fig:sigma} are underestimated with respect to the 1000 reference Patchy or LN mocks by up to $\sim40\%$, for both monopole and quadrupole. This effect is mitigated when the number of jackknife resamplings increases.
 
 Beyond BAO scales, we observe the opposite trend, with CMASS jackknife errors tendentially overestimated with respect to the ideal cases. Around 130\,$h^{-1}$Mpc, the monopole errors from CMASS covariances are $\sim10-50\%$ larger than those from 1000 Patchy or LN mocks and the discrepancy overall increases with the number of resamplings. Compared to the monopole, the quadrupole shows smaller fluctuations in the 1$\sigma$ ratio shown in the top panels of Fig.\,\ref{fig:sigma}.

 The amplitude of the quadrupole 20\, JK result is $\sim10-20\%$ lower than the others on scales below 150\,$h^{-1}$Mpc; the monopole is lower than the rest on about all scales when compared to the 1000 Patchy mocks, and only beyond 140\,$h^{-1}$Mpc when compared to the 1000 LN. 
As expected, the 20 and 50 jackknife schemes return the largest fluctuations. Although larger jackknife regions with greater independence may give a more accurate covariance estimate, the uncertainty on the covariance is large due to having few resamplings available. In the 20 JK scheme coupled with 10 $s$ bins, the large fluctuations produced by a limited number of resamplings are partially mitigated by the smaller number of bins compared to the other cases.

In the bottom panels of Fig.\,\ref{fig:sigma}, we display the ratios of the average variance from jackknife resamplings on 10 Patchy or LN realisations and that from the 1000 independent mocks of reference. The Patchy JK outcomes show similar fluctuations and trends to the CMASS results shown in the top left panel, with deviations from the reference uncertainties of up to 40-50\% on large scales.

On the other hand, the LN jackknife ratios exhibit flatter trends, similar to the CMASS results in the top right panel. The biggest difference is seen in the LN JK monopole signals, which are overestimated (underestimated) on small (large) scales by up to $\sim50\%$ ($\sim40\%$). Beyond $\sim150\,h^{-1}$Mpc, the monopole (quadrupole) errors from LN jackknife seems to converge to the results from 1000 LN independent mocks. Again, the 20 and 50 JK configurations are the ones exhibiting the largest fluctuations due to the limited number of jackknife resamplings. 

Overall, we find reasonable agreement between the uncertainties from jackknife on CMASS, Patchy or LN mocks with the same volume of CMASS, and from 1000 independent Patchy/LN realisations. In general, with respect to the 1000 reference mocks representing the ideal case, the monopole errors from CMASS JK are underestimated on small scales and overestimated beyond $\sim80\,h^{-1}$Mpc, those from Patchy JK are overestimated on all scales, and those from LN JK are overestimated below $\sim80\,h^{-1}$Mpc and underestimated beyond.
For the quadrupole, the CMASS JK errors are overestimated (underestimated) compared to the 1000 Patchy (LN) mocks of reference, while those from Patchy (LN) jackknife are overestimated compared to both 1000 Patchy and LN mocks. All these discrepancies remain within 30\% in most cases.

Despite the use of only 10 mock realisation is not sufficient to clearly identify general trends, the jackknife variances in Fig.\,\ref{fig:sigma} reveal deviations from that of the full set of mocks. For Patchy, these differences increase on larger scales in both multipoles, while for the LN mocks they peak on smaller and intermediate scales. These deviations are due to the known limitations of the jackknife method which, nevertheless, provides covariance estimates in reasonable agreement with those from multiple mock realisations.

\begin{figure}
\begin{center}\vspace{-0.3cm}
\includegraphics[width=\linewidth]{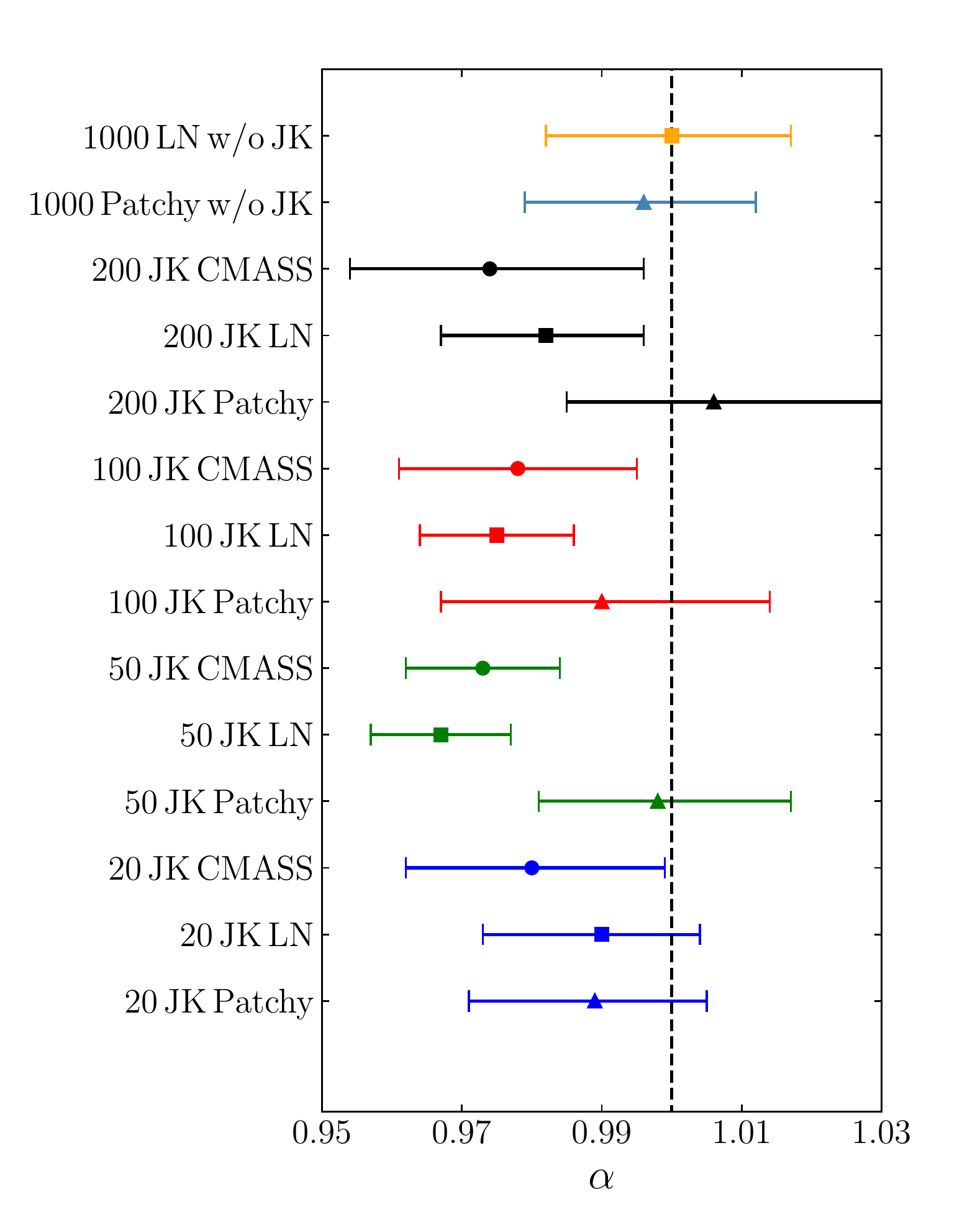}\vspace{-0.6cm}
\caption{Summary of the $\alpha$ shift parameters obtained from the covariances calculated either using the 1000 independent mock realisations, or the JK configurations and binning schemes reported in Table \ref{tab:results_params}. The results from JK on LN or Patchy are the average values of 10 $\alpha$ estimates, each one obtained from a different mock realisation. The points are color-coded as in Fig.\,\ref{fig:2PCF}, where each colour corresponds to a different jackknife/binning scheme. The results from CMASS are represented by dots, those from LN by squares and those for Patchy mocks by triangles. The vertical line shows the value $\alpha=1$ to help the comparison. 
All these results are calculated assuming a tapering parameter $T_{\rm{p}}=500$. }
\label{fig:alpha_plot}
\end{center}
\end{figure}

We next analyse the impact of the different covariance estimates on the error of the $\alpha$ shift parameter.
Fig.\,\ref{fig:alpha_plot} compares the $\alpha$ values and corresponding uncertainties  obtained from the covariances based on (i) BOSS CMASS JK, (ii) Patchy/LN JK, and (iii) 1000 independent Patchy/LN mocks. For each jackknife configuration performed on the LN or Patchy mocks, we show the average $\alpha$ value from 10 different realisations randomly chosen.  In general, a sample size of 10 is sufficient to infer the mean and scatter with a precision of $\sim 30$\%. Compared with just a single mock, in this way we improve the precision of the error estimate by $\sqrt{10}$. 
These results are reported in Tab.\,\ref{tab:results_params} and assume a tapering parameter $T_{\rm{p}}=500$, which we found to be optimal. 

Overall, we find good agreement between the results based on covariances from jackknife, both applied to CMASS observations, LN and Patchy mocks. 
The 1000 LN and Patchy results without jackknife are consistent with the JK outcomes, despite the difference in the pre-factor of their covariances (Eqs.\,\ref{eq:covma},\ref{eq:covmocks}) and the difference in the structure of their off-diagonal terms (see Fig.\,\ref{fig:matrixmock}). 

The uncertainties on $\alpha$ are all in agreement with each other, independently from the number/size of jackknife resamplings adopted. The average errors obtained from 200, 100 and 50\,JK resamplings (i.e. the most robust ones) performed on CMASS data, 10 LN and 10 Patchy mocks are $\sim1.6\%$, $\sim1.2\%$ and $2.2\%$, respectively. Those from the 1000 LN and Patchy covariances without jackknife are $\sim1.7\%$ and $\sim1.6\%$, respectively. These last two cases are ideal since the 1000 LN and Patchy mocks are all independent (but we do not expect the log-normal mocks to capture the full covariance of the CMASS galaxy sample).
The 50 and 20\,JK schemes are the ones returning the largest fluctuations in the covariances, which can result in errors on $\alpha$ as large as $\sim1.9\%$. These are not the largest uncertainties on $\alpha$, which instead come from the 200 JK and 100 JK Patchy configurations. 
In order to quantify the fluctuations, we have repeated all the jackknife resamplings on 10 different realisations of both the LN and Patchy mocks.

The constraints on $\alpha$ presented in Fig.\,\ref{fig:alpha_plot} do not show a clear trend with $N_{\rm JK}$, and highlight a different behaviour between the Patchy and the LN mocks, which seems to converge for the 20\,JK case.

\begin{figure}
\begin{center}\vspace{-0.2cm}
\includegraphics[width=\linewidth]{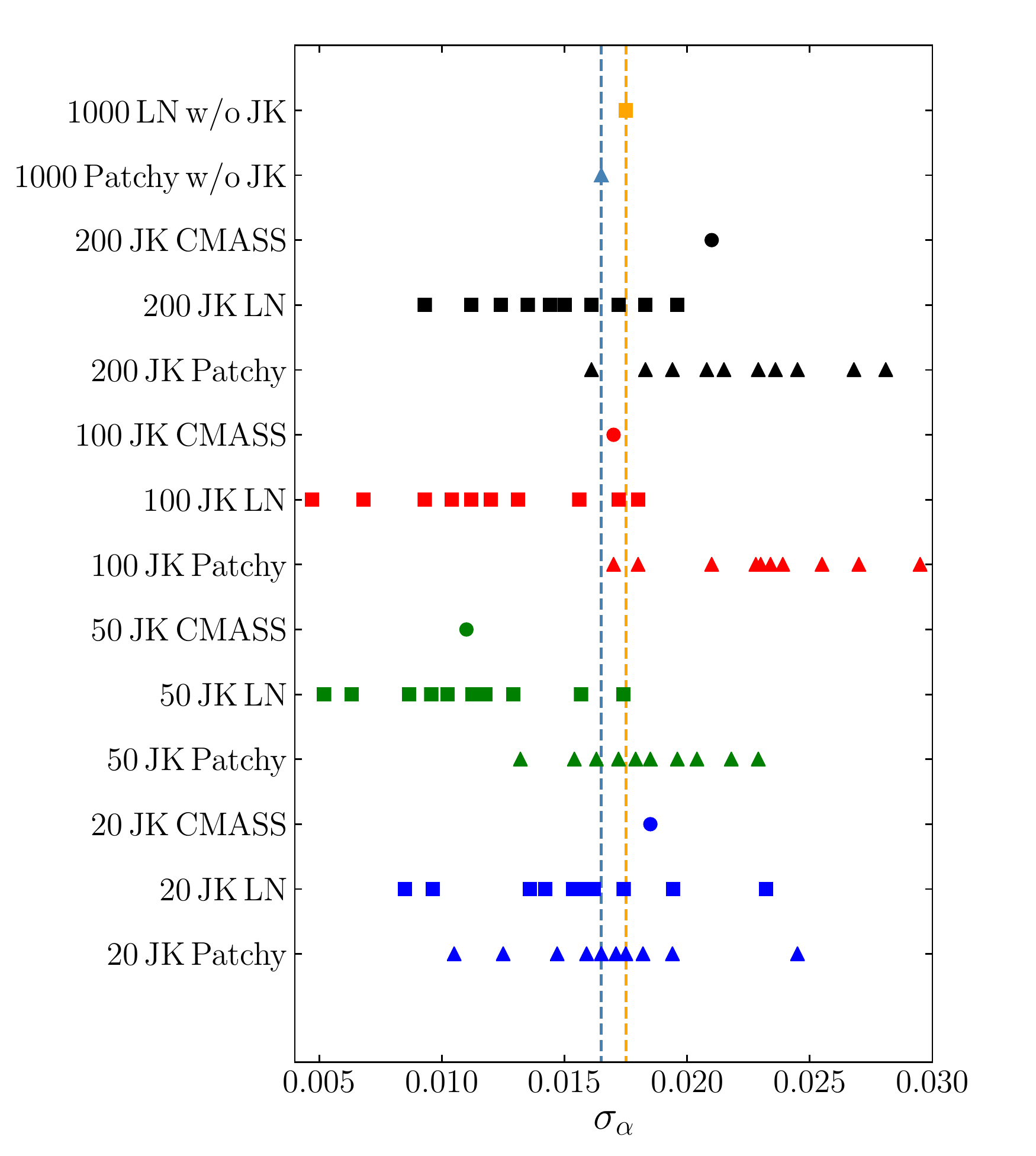}\vspace{-0.6cm}
\caption{Uncertainties on the $\alpha$ shift parameter obtained from the 1000 Patchy and LN mocks without jackknife, and from the CMASS, 10 Patchy and 10 LN jackknife resamplings. The different sets of points are color-coded as in Fig.\,\ref{fig:alpha_plot} and the vertical lines help the comparison with the ideal cases of 1000 Patchy and LN mocks.}
\label{fig:sigma_alpha_plot}
\end{center}
\end{figure}

In Fig.\,\ref{fig:sigma_alpha_plot}, we present all the uncertainties on $\alpha$ computed in our analysis. Each point represents the mean of the corresponding upper and lower $\sigma$ values. Besides the ideal cases of 1000 Patchy and LN mocks, we show the errors from CMASS jackknife and those from jackknife on 10 different Patchy and LN realisations. The errors are all consistent and we cannot identify a specific jackknife configuration that returns a better estimate of $\sigma_{\alpha}$. The Patchy and LN mocks are different, but they converge in the 20 JK configuration.

It is noteworthy to mention that the alpha results in Fig.\,\ref{fig:sigma_alpha_plot} include the Hartlap factor and demonstrate that this correction is needed for jackknife covariances.  In fact, if we do not include it, the error on the 20\,JK case would be smaller by $\sim \sqrt{0.42}$ (see Table\,\ref{tab:hartlap}). In conclusion, only when the Hartlap factor is applied all the jackknife configurations give a consistent error on $\alpha$.

\begin{table*}
\begin{center}
\begin{tabular}{lcccc}
\hline
\multicolumn{3}{r|}{\hspace{2.5cm}$\alpha$}\\
&BOSS CMASS & LN light-cones& Patchy mocks\\
\hline
200 JK, 20 bins:&$0.974^{+0.022}_{-0.020}$ & $0.982^{+0.015}_{-0.014}$&$1.006^{+0.021}_{-0.025}$\\\\
100 JK, 20 bins:&$0.978^{+0.017}_{-0.017}$ & $0.975^{+0.011}_{-0.011}$&$0.990^{+0.023}_{-0.024}$ \\\\
50 JK, 20 bins: &$0.973^{+0.011}_{-0.011}$ & $0.967^{+0.010}_{-0.010}$&$0.998^{+0.017}_{-0.019}$ \\\\
20 JK, 10 bins: &$0.980^{+0.019}_{-0.018}$ & $0.990^{+0.017}_{-0.014}$&$0.989^{+0.018}_{-0.016}$ \\\\
\hline
All LN (Patchy) w/o JK, 20 bins:& &$1.000^{+0.018}_{-0.017}$&$0.996^{+0.017}_{-0.016}$ \\
\hline 
\end{tabular}
\end{center}
\caption{Estimates of the $\alpha$ shift parameter and its uncertainty obtained from the four jackknife configurations coupled with two binning schemes applied to the CMASS data, the log-normal lightcones and the Patchy mocks. For the LN and Patchy mocks we show the mean $\alpha$ value obtained from the jackknife runs on 10 realisations randomly chosen. The last row shows the result obtained from the covariances of the 1000 LN or Patchy mocks without performing jackknife resampling. All these results here assume an optimal tapering parameter of $T_{\rm{p}}=500$ and are shown in Fig.\,\ref{fig:alpha_plot}.} 
\label{tab:results_params}
\end{table*}
\begin{figure}
\begin{center}
\includegraphics[width=\linewidth]{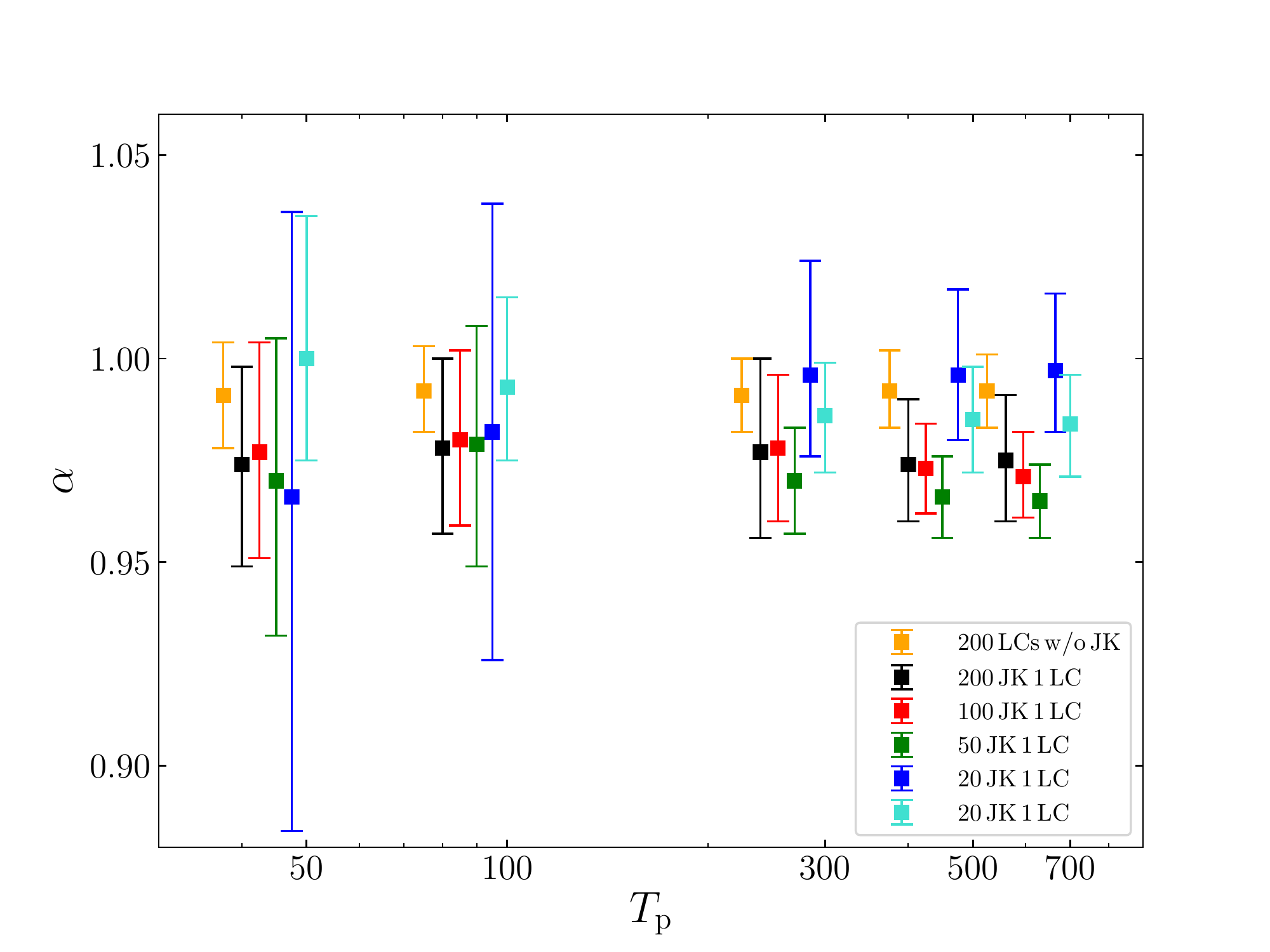}
\includegraphics[width=\linewidth]{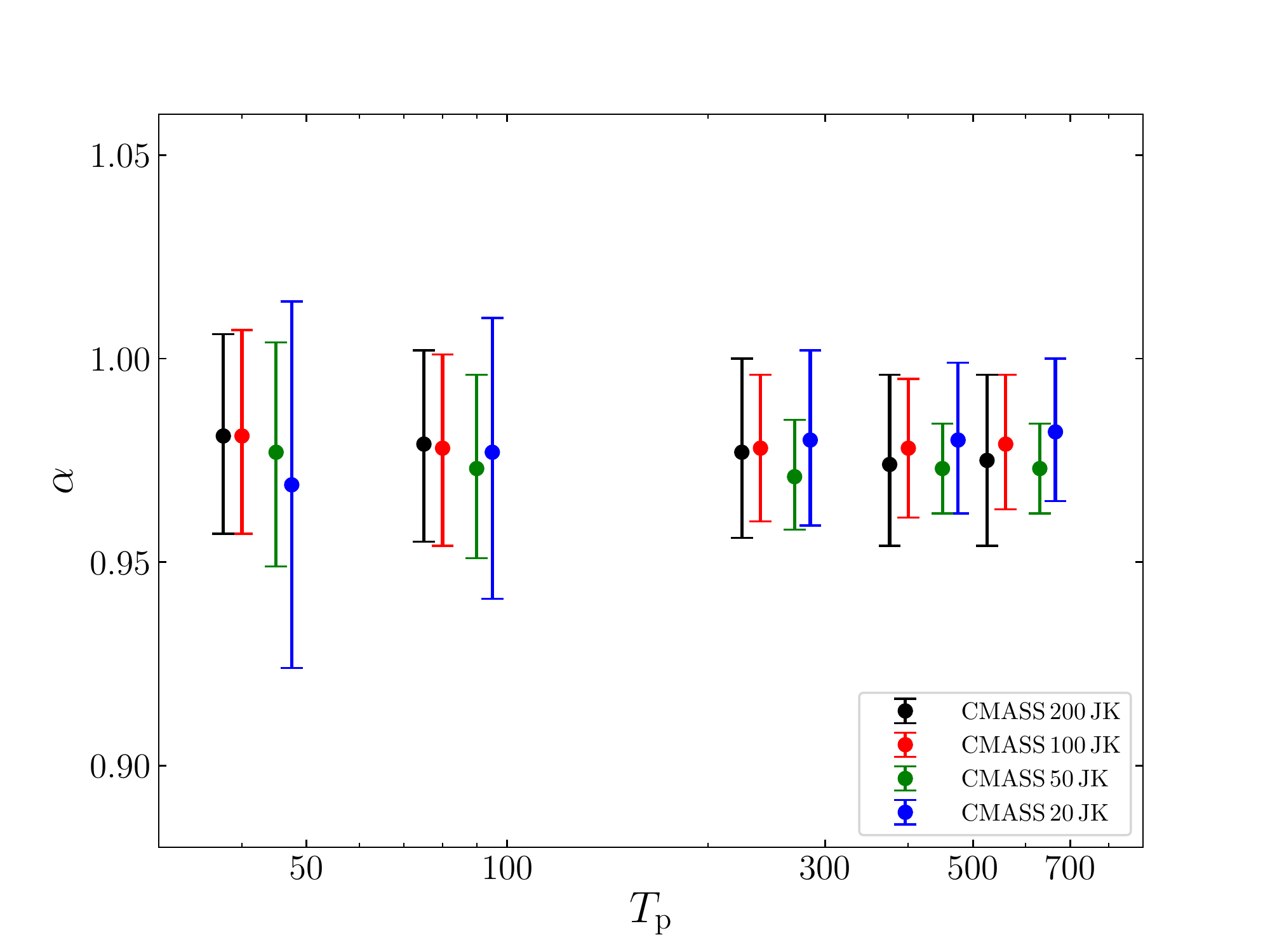}\vspace{-0.5cm}
\caption{Shift parameter $\alpha$ and its uncertainty as a function of the tapering parameter $T_{\rm{p}}$. \textit{Top:} results from covariances computed from 200 LN mocks without jackknife and from jackknife performed on a LN realisation. We have offset the $T_{\rm{p}}$ values on the x-axis by multiplying them, from left to right, by $[0.75,0.80,0.85,0.90,0.95,1.0]$. \textit{Bottom:} results from CMASS jackknife resampling. The $T_{\rm{p}}$ values have been offset by multiplying them, from left to right, by $[0.80,0.85,0.90,0.95,1.0]$.}
\label{fig:alphaT}
\end{center}
\end{figure}

We have explored further the dependence of the $\alpha$ parameter and its uncertainty on the tapering parameter $T_{\rm p}$. In Fig.\,\ref{fig:alphaT} we show the $\alpha$ results assuming $T_{\rm{p}}=[50,\,100,\,300,\,500,\,700]$. The top panel displays $\alpha$ from the 1000 LN mocks without jackknife and from jackknife applied to one of the LN; the bottom panel shows the CMASS jackknife outcomes. In the plots we offset the tapering values by a multiplicative factor to avoid crowding (see caption of Fig.\,\ref{fig:alphaT}).
The optimal value, which provides errors on $\alpha$ of $\sim1-2\%$, turns out to be $T_{\rm{p}}=500$. In this way, our BOSS CMASS $\alpha$ estimates are comparable with previous results in the literature \citep[pre-reconstruction, e.g.][]{10.1093/mnras/stw2372,10.1093/mnras/sty1266}.

The 20\,JK configuration shows the largest fluctuations due to the limited number of resamplings. 
We have further tested our MCMC code without including any tapering correction and leaving only the Hartlap factor. In this case we find that the covariances from 20\,JK resamplings are no longer semi-positive definite, meaning that they are not invertible, hence not useful for assembling the precision matrix needed to estimate $\alpha$. Such a result confirms that jackknife configurations with few cells tend to provide non robust covariance estimates.

\section{Discussion and summary}
\label{sec:discussion}
We have studied the impact of choosing different sizes and numbers of jackknife resamplings on the accuracy of the covariance estimates and the $\alpha$ shift parameter.
To this purpose, we have measured the first two even multipoles of the BOSS CMASS DR12 galaxy sample at $0.43<z<0.7$ and we have modelled the results both using 1000 MultiDark Patchy mocks (Sec.\,\ref{sec:patchy}), a set of 1000 log-normal mocks (Sec.\,\ref{sec:mocks}) and an analytic approach (Sec.\,\ref{sec:analyticmodel}). We have computed their covariances using 200, 100, 50 and 20 jackknife resamplings coupled with two binning schemes: 20 or 10 linear bins in $24<s<184\,h^{-1}$Mpc, with 120 linear bins in $0<\mu<1$ (see Sec.\,\ref{sec:measurements}). We have compared the results with the covariances obtained from the 1000 independent Patchy and LN mocks without jackknife. We have then applied the same jackknife configurations above on 10 different Patchy/LN realisations to derive covariances directly comparable with the CMASS ones. 

From these different covariance matrices we have derived corresponding precision matrices (Sec.\,\ref{sec:javiermethod}), which we have used as inputs for our Monte Carlo Markov Chain to estimate the baryon acoustic scale through the $\alpha$ shift parameter and its uncertainty. Our main findings are summarised in what follows:
\begin{itemize}
    \item We find good consistency between the covariances obtained from CMASS, Patchy and LN jackknife resamplings, and from 1000 Patchy/LN without jackknife.  This leads to consistent error bars in both the galaxy clustering measurements and the $\alpha$ shift parameter.
    \item We find no evidence for a bias in the inferred value of $\alpha$ or its error when the jackknife cell size is smaller than the maximum 2PCF scale measured. However, with few resamplings available the error estimate becomes unreliable. This result may be due to the fact that the JK volumes in our analysis have a dimension along the line-of-sight that is longer than the maximum pair separation. Thus, even though we find no effect on the isotropic $\alpha$ parameter, there may be a bias if we separate the  LOS and transverse modes by fitting $\alpha_{\parallel}$ and $\alpha_{\bot}$.
    \item We have demonstrated that it is useful to apply the Hartlap factor and the tapering scheme to estimate the precision matrix with jackknife resampling. The $\alpha$ shift parameter estimated either from CMASS, Patchy or LN jackknife covariances, or from 1000 independent Patchy or LN  mocks without jackknife, are all consistent between each other and in reasonable agreement with previous BOSS CMASS DR12 results from galaxy clustering pre-reconstruction analysis \citep{10.1093/mnras/stw2372}. We find uncertainties on $\alpha$ of 1-2.5\%, depending on the jackknife size and 2PCF binning scheme adopted. This confirms that the jackknife methodology applied to both observations and mocks produces a comparable level of noise in the covariance estimates. This noise is then reduced in the precision matrix by applying the tapering correction (see Sec.\,\ref{sec:javiermethod}).
    \item We have tested different values for the tapering parameter, in the range $50\leq T_{\rm{p}}\leq 700$, to maximise the accuracy in the $\alpha$ shift parameter estimation. We find that the optimal value is $T_{\rm{p}}=500$. By lowering it, the noise in the precision matrix estimate is suppressed but the error on $\alpha$ grows.
\end{itemize}

To summarise, performing jackknife resamplings either on BOSS CMASS DR12 data, Patchy or log-normal mocks with the same CMASS volume provides covariances that are consistent with those obtained from a set of 1000 independent Patchy or LN mocks and with previous results in the literature \citep{10.1093/mnras/stw2372}. These covariances lead to $\alpha$ estimates with 1-2.5\% uncertainties, depending on the jackknife size/2PCF binning scheme assumed.

The largest differences between covariance estimates from jackknife resampling and 1000 independent mock realisations without JK are visible in the off-diagonal terms. Here the jackknife results exhibit a higher level of noise. This difference is key for determining the accuracy of the $\alpha$ shift parameter. The action of the tapering correction (Sec.\,\ref{sec:javiermethod}) is to considerably reduce this noise returning comparable uncertainties on $\alpha$ from all of the different covariance estimates tested.

Although previous works limit the jackknife scale to larger than the measured 2PCF scale \citep[e.g.][]{Beutler2011,2016ApJ...826..154H}, we find that this is not essential. In fact, when using jackknife to estimate covariances, one should prioritise building a large number of resamplings rather than choosing a jackknife size larger than the maximum galaxy clustering scale measured. In fact, especially when studying BAO scales, by requiring $S_{\rm{JK}}\ge{\rm{max}}(s)$, we are able to build only few wide jackknife regions, which leads to large uncertainties in the error estimates.
In our results, we do not see a clear trend that the $\alpha$ error bars tend to reduce with a specific jackknife configuration. All the JK configurations tested return consistent errors.

The new generation of cosmological surveys, such as DESI, Euclid or LSST, will span larger volumes compared to SDSS-III/BOSS. The precision in their covariance estimates based on thousands of independent mock realisations will depend on the feasibility (and our ability) of producing large synthetic data sets. Whenever mocks will not be available, the jackknife method can be used to obtain reasonable covariance estimates. The precision of the cosmological parameters inferred from the jackknife covariances will be determined in part by the number of resamplings. 

We find that it is not essential to use jackknife sizes larger than the BAO scale, and so it will be possible to achieve $N>10^3$ resamplings to reach percent level precision on the error of cosmological parameters using the jackknife approach. From our analysis we have found that the covariance estimates are less noisy when a large number of jackknife resamplings is performed. However, the noise in the covariances has relatively small effect on the $\alpha$ uncertainty. In a followup work, we will address the feasibility of inferring covariance estimates for a survey such as Euclid using a large number of jackknife resamplings.

\section*{Acknowledgements}
GF acknowledges financial support from the SNF 175751 “Cosmology with 3D Maps of the Universe” research grant. 

GF thanks the Insitute of Cosmology and Gravitation at Portsmouth University to host and fund her through a Dennis Sciama fellowship during the first part of this project.

DS acknowledges financial support from the Fondecyt Regular project number 1200171. 

The authors are thankful to Cheng Zhao, who provided the 1000 Patchy pair counts, allowing them to save a considerable amount of time.

Funding for SDSS-III has been provided by the Alfred P. Sloan Foundation, the Participating Institutions, the National Science Foundation, and the U.S. Department of Energy Office of Science. The SDSS-III web site is \url{http://www.sdss3.org/}.

SDSS-III is managed by the Astrophysical Research Consortium for the Participating Institutions of the SDSS-III Collaboration including the University of Arizona, the Brazilian Participation Group, Brookhaven National Laboratory, Carnegie Mellon University, University of Florida, the French Participation Group, the German Participation Group, Harvard University, the Instituto de Astrofisica de Canarias, the Michigan State/Notre Dame/JINA Participation Group, Johns Hopkins University, Lawrence Berkeley National Laboratory, Max Planck Institute for Astrophysics, Max Planck Institute for Extraterrestrial Physics, New Mexico State University, New York University, Ohio State University, Pennsylvania State University, University of Portsmouth, Princeton University, the Spanish Participation Group, University of Tokyo, University of Utah, Vanderbilt University, University of Virginia, University of Washington, and Yale University.

\section{Data availability}
The BOSS DR12 CMASS data analysed in this work are publicly available at \url{https://dr12.sdss.org/sas/dr12/boss/lss/}, The 1000 Patchy mocks can be downloaded at \url{https://data.sdss.org/sas/dr12/boss/lss/dr12\_multidark\_patchy_mocks/}. The 1000 log-normal mocks have been generated using \textit{Synmock} and they are available upon request to the authors.



\bibliographystyle{mnras}
\bibliography{references}

\bsp	
\label{lastpage}
\end{document}